\documentstyle[fleqn,epsfig,run2col]{article}

\def\D0{D\O}
\def\ETmiss{{\rm {\mbox{$E\kern-0.57em\raise0.19ex\hbox{/}_{T}$}}}}

\def\met{\mbox{${\hbox{$E$\kern-0.6em\lower
               -.1ex\hbox{/}}}_{T}$}} 
\def\mpt{\mbox{${\hbox{$p$\kern-0.6em\lower
               -.05ex\hbox{/}}}_{T}$}} 
\def\mht{\mbox{${\hbox{$H$\kern
               -0.6em\lower-.1ex\hbox{/}}}_{T}$}} 
\def\vmet{\mbox{${\hbox{$\vec{E}$\kern
                -0.6em\lower-.1ex\hbox{/}}}_{T}$}}
\def\vmpt{\mbox{${\hbox{$\vec{p}$\kern
                -0.4em\lower-.1ex\hbox{/}}}_{T}$}}
\def\lsim{\mathrel{\rlap{\lower4pt\hbox{\hskip1pt$\sim$}}
    \raise1pt\hbox{$<$}}}         
\def\gsim{\mathrel{\rlap{\lower4pt\hbox{\hskip1pt$\sim$}}
    \raise1pt\hbox{$>$}}}         

\newcommand{\AmS}{{\protect\the\textfont2
  A\kern-.1667em\lower.5ex\hbox{M}\kern-.125emS}}
\def\KT{\mbox{$K_T$} }
\def\ppbar{\mbox{$p\overline{p}$}}
\def\gev{GeV}
\def\pt{\mbox{$p_{T}$}}

\title{Run II Jet Physics}

\author{Gerald C. Blazey\address{Department of Physics, Northern Illinois
        University, DeKalb, IL 60115, USA},
        Jay R. Dittmann\address{Fermilab, P.O. Box 500, Batavia, IL 60510, USA},
        Stephen D. Ellis\address{Department of Physics, University of Washington,
        Box 351560, Seattle, WA 98195-1560, USA},
        V. Daniel Elvira$^{\rm b}$,
        K.~Frame\address{Department of Physics and Astronomy, Michigan State University,
        East Lansing, MI 48824, USA},
        S.~Grinstein\address{Depto. de Fisica, FCEyN-Universidad de Buenos Aires,
        Pab I, Ciudad Universitaria, (1428) Capital Federal, Argentina},
        Robert Hirosky\address{Department of Physics, University of Illinois at
        Chicago, Chicago, IL 60607, USA},
        R.~Piegaia$^{\rm e}$,
        H.~Schellman\address{Physics Department, Northwestern University,
        Evanston, IL 60210},
        R.~Snihur$^{\rm g}$,
        V.~Sorin$^{\rm e}$,
        Dieter Zeppenfeld\address{Department of Physics, University of Wisconsin at
        Madison, Madison, WI 53706, USA}
        }

\begin{document}

\begin{abstract}
The Run II jet physics group includes the Jet Algorithms, Jet
Shape/Energy Flow, and Jet Measurements/Correlations subgroups. The
main goal of the jet algorithm subgroup was to explore and define
standard Run II jet finding procedures for CDF and D\O. The focus
of the jet shape/energy flow group was the study of jets as objects
and the energy flows around these objects. The jet
measurements/correlations subgroup discussed measurements at
different beam energies; $\alpha _{S}$ measurements; and LO, NLO,
NNLO, and threshold jet calculations. As a practical matter the
algorithm and shape/energy flow groups merged to concentrate on the
development of Run II jet algorithms that are both free of theoretical
and experimental difficulties and able to reproduce Run I measurements.

Starting from a review of the experience gained
during Run I, the group considered a variety of cone algorithms and \KT
algorithms.  The current understanding of both types
of algorithms, including calibration issues, are discussed in this report
along with some preliminary experimental results.
The jet algorithms group recommends that CDF and D\O\ employ the {\it same} version
of {\it both} a cone algorithm and a \KT
algorithm during Run II.  Proposed versions of each type of
algorithm are discussed.  The group also recommends the use of
full 4-vector kinematic variables whenever possible.
The recommended algorithms attempt to minimize the impact of seeds in the case of
the cone algorithm
and preclustering in the case of the \KT algorithm.
Issues regarding precluster definitions and merge/split criteria require further
study.

\end{abstract}

\maketitle

\section{Prologue}

The Run I jet programs at CDF and D\O\ made impressive measurements of the
inclusive jet cross section, dijet angular and mass distributions, and
triple differential cross sections. These measurements were all marked by
statistical accuracy equal or superior to current theoretical
accuracy~\cite{AREV}. However, the always compelling search for quark
compositeness, the
quest to improve the calculational accuracy of QCD, and the desire to fully
understand the composition of the proton will certainly prompt improvements
over these measurements. Without question, with $\sim $2 \mbox{$fb^{-1}$},
the Run II jet physics program will extend the jet measurements of Run I to
even higher jet energies.

There are three issues, experimental and theoretical, that
currently limit the sensitivity of compositeness searches and QCD
tests: limited knowledge of the parton distribution functions
(pdfs), systematic uncertainties related to jet energy calibration,
and the limited accuracy of fixed order perturbative calculations
due to the incomplete nature of the calculations and incomplete
specification of jet finding algorithms. Inadequate knowledge of
the pdfs and calibration are currently the dominant uncertainties,
engendering greater than 50\% uncertainties at the largest
energies. The reader may refer to the chapter on Parton
Distributions for a complete discussion of pdf measurements.

As mentioned, the uncertainty of NLO perturbative calculations is
due in part to the inherent incompleteness of fixed order
calculations. The initial meeting of the jet physics group included
talks on ``Leading Order (LO) Multi-jet Calculations'' by
Michelangelo Mangano, ``Next-to-Leading Order (NLO) Multi-jet
Calculations'' by Bill Kilgore, ``Prospects for Next-to-NLO (NNLO)
Multi-jet Calculations'' by Lance Dixon, ``Threshold Resummations
for Jet Production'' by Nicolas Kidonakis, ``Different Beam
Energies'' by Greg Snow, and ``$\alpha _{S}$ Measurements in Jet
Systems'' by Christina Mesropian. These attempts to improve the
accuracy of perturbative calculations show the vigorous nature of
ongoing efforts and should prove fruitful before the arrival of Run
II data.

Jet algorithms, the other source of calculation uncertainty, start
from a list of ``particles'' that we take to be calorimeter towers
or hadrons at the experimental level, and partons in a perturbative
QCD calculation. \ The role of the algorithm is to associate
clusters of these particles into jets such that the kinematic
properties of the jets ({\it e.g}., momenta) can be related to the
corresponding properties of the energetic partons produced in the
hard scattering process. \ Thus the jet algorithm allows
us to ``see'' the partons (or at least their fingerprints) in the
hadronic final state. \

Differences in the properties of reconstructed jets when
going from the parton to the hadron or calorimeter level are a
major concern for a good jet algorithm. Each particle $i$ carries a
4-momentum $p_{i}^{\mu }$, which we take to be massless. The
algorithm selects a set of particles, which are typically emitted
close to each other in angle, and combines their momenta to form
the momentum of a jet. The selection process is called the ``jet
algorithm'' and the momentum addition rule is called the
``recombination scheme''. \ Note that these two steps are logically
distinct. \ One can, for example, use one set of kinematic
variables in the jet algorithm to determine the particles in a jet
and then construct a separate set of kinematic variables to
characterize the jets that have been identified. \ This point will
be important in subsequent discussions.

Historically cone algorithms have been the jet algorithm of choice
for hadron-hadron experiments. As envisioned in the Snowmass
algorithm~\cite{SNOWMASS}, a cone jet of radius $R$ consists of all
of the particles whose trajectories (assuming no bending by the
magnetic field of the detector) lie in an area $A = \pi R^2$ of
$\eta \times \phi$ space, where $\eta$ is the pseudorapidity
$\eta=-\ln\tan \theta/2$.  It is further required, as explained in
detail below, that the axis of the cone coincides with the jet
direction as defined by the $E_T$-weighted centroid of the
particles within the cone (where $E_T$ is transverse energy,
$E_T=E\sin\theta$). In
principle, one simply searches for all such ``stable'' cones to
define the jet content of a given event.

In practice, in order to save computing time, the iterative process
of searching for the ``stable'' cones in experimental data starts
with only those cones centered about the most energetic particles
in the event (the so-called ``seeds''). Usually, the seeds are
required to pass a threshold energy of a few hundred MeV in order
to minimize computing time. The $E_T$-weighted centroids are
calculated for the particles in each seed cone and then the
centroids are used as centers for new cones in $\eta
\times \phi$ space. This procedure is iterated for each cone until
the cone axis coincides with the centroid.
Unfortunately, nothing prevents the final stable cones from overlapping.
A single particle may belong to two or
more cones.  As a result, a procedure must be included in the
cone algorithm to specify how to split or merge overlapping
cones~\cite{rsep}.

At least part of the uncertainty associated with fixed order
perturbative calculations of jet cross sections can be attributed to the
difficulties encountered when this experimental jet cone algorithm,
with both seeds and merging/splitting rules, is applied to
theoretical calculations. (See Ref.~\cite{AREV} for a discussion of the
CDF and \D0 algorithms.)  Neither issue was treated by the original
Snowmass algorithm~\cite{SNOWMASS} that forms the basis of
fixed order perturbative cone jet calculations.  Current
NLO inclusive jet cross section
calculations (which describe either two or three final state
partons) require the addition of an ad hoc parameter
$R_{Sep}$ ~\cite{RSEP}. This additional parameter is used to regulate
the clustering of partons and simulate the role of seeds and
merging in the experimentally applied algorithm. In essence, the
jet cone algorithm, used so pervasively at hadron-hadron colliders,
must be modeled in NLO calculations. This modeling results in
2--5\% uncertainties as a function of jet transverse energy $E_{T}$
in calculated cross sections.

Even worse, with the current cone algorithms, cross sections
calculated at NNLO exhibit a marked sensitivity to soft radiation.
As an illustration, consider two well-separated partons that will
just fit inside, but at opposite sides, of a single cone.  With
only the two partons, and nothing in between to serve as a seed,
the current standard cone algorithms will reconstruct the two
partons as two jets.  At NNLO a very soft gluon could be radiated
between the two well-separated partons and serve as a seed.  In
this case the single jet solution, with both partons inside, will
be identified by the current cone algorithm. Thus the outcome of
the current cone algorithm with seeds is manifestly sensitive to
soft radiation. Because of the difficulties inherent with typical
usage of the cone algorithm, the jet algorithm and jet shape/energy
flow subgroups decided to establish an Improved Legacy Cone
Algorithm (whimsically dubbed ILCA).  Ideally, the ILCA should
replicate Run I cross sections within a few percent, but not have
the same theoretical difficulties.

Inspired by QCD, a second class of jet algorithms, \KT algorithms,
has been developed.  These algorithms successively merge
pairs of ``particles'' in order of increasing relative transverse
momentum. They typically contain a parameter, $D$ (also
called $R$), that controls termination of merging and characterizes
the approximate size of the resulting jets. Since a \KT algorithm
fundamentally merges nearby particles, there is a close
correspondence of jets reconstructed in a calorimeter to jets
reconstructed from individual hadrons, leptons and photons.
Furthermore, every particle in an event is assigned to a unique
jet. Most importantly, \KT jet algorithms are, by design, infrared
and collinear safe to all orders of calculation. The algorithms can
be applied in a straightforward way to fixed--order or resummed
calculations in QCD, partons or particles from a Monte Carlo event
generator, or energy deposited in a detector~\cite{RECOMB}.

However, until recently, a full program for the calibration of \KT
algorithms at hadron-hadron colliders had not been developed.
This was due mostly to difficulties with the subtraction of
energy from spectator fragments and from the pile-up of multiple
hadron-hadron interactions. Since the \KT jets have no fixed
shape, prescriptions for dealing with the extra energy have been
difficult to devise and the use of \KT algorithms at
hadron-hadron colliders has been limited.  Also, as with the issue
of seeds in the case of
the cone algorithm, there is a practical question of minimizing
the computing time required to apply the \KT algorithm.  Typically
this is treated in a preclustering step where the number of ``particles''
is significantly reduced before the \KT algorithm is applied.
A successful \KT algorithm must ensure that any preclustering step
does not introduce the sort of extra difficulty found with seeds.

Buoyed by the successful use of \KT algorithms at LEP and HERA,
eager to benefit from their theoretical preciseness, and
reassured by recent success with calibration, the jet physics group
decided to specify a standard \KT algorithm for Run II.

\section{Attributes of the Ideal Algorithm}

Although it provided a good start, the Snowmass algorithm has
proved to be incomplete. It does not address either the phenomena
of merging and splitting or the role of the seed towers with the
related soft gluon sensitivity. Also, jet energy and angle
definitions have varied between experiments. To treat these issues,
the group began discussions with the following four general
criteria:

\begin{enumerate}
\item  {\em Fully Specified}: The jet selection process, the jet
kinematic variables and the various corrections ({\it e.g.},
the role of the underlying
event) should be clearly and completely defined. If necessary,
preclustering, merging, and splitting algorithms must be
completely described.

\item  {\em Theoretically Well Behaved}: The algorithm should be
infrared and collinear safe with no ad hoc clustering parameters.

\item  {\em Detector Independence}: There should be no dependence on cell
type, numbers, or size.

\item  {\em Order Independence}: The algorithms should behave equally at
the parton, particle, and detector levels.
\end{enumerate}

The first two criteria should be satisfied by every algorithm;
however, the last two can probably never be exactly true, but
should be approximately correct.

\subsection{Theoretical Attributes of the Ideal \\ Algorithm}

\begin{figure}[htb]
\begin{center}
\epsfig{figure=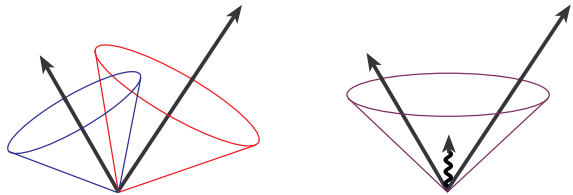,width = 75mm}
\end{center}
\caption{An illustration of infrared sensitivity in cone jet clustering. In this
example, jet clustering begins around seed particles, shown here as
arrows with length proportional to energy. We illustrate how the
presence of soft radiation between two jets may cause a merging of
the jets that would not occur in the absence of the soft
radiation.}
\label{IRSafe}
\end{figure}

\begin{figure}[tbp]
\begin{center}
\epsfig{figure=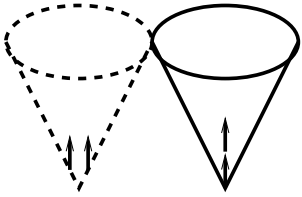,width=40mm}
\end{center}
\caption{An illustration of collinear sensitivity in jet reconstruction.
In this example, the configuration on the left fails to produce a seed
because its energy is split among several detector towers. The
configuration on the right produces a seed because its energy is
more narrowly distributed.}
\label{COsafe}
\end{figure}

\begin{figure}[ht]
\begin{center}
\epsfig{figure=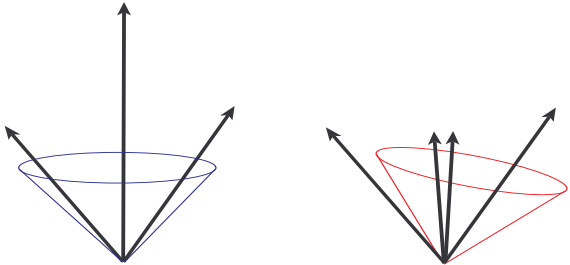,height=35mm,width=75mm}
\end{center}
\caption{Another collinear problem. In this case we illustrate
possible sensitivity to $E_T$ ordering of the particles that act
as seeds.}
\label{COsafe2}
\end{figure}

The initial efforts of the algorithm working group were focused on
extending and illuminating the list of desirable features of an
``ideal'' jet algorithm. From the ``theoretical standpoint'' the
following features are desirable and, for the most part, necessary:

\begin{enumerate}
\item  {\em Infrared safety:} The algorithm should not only be infrared
safe, in
the sense that any infrared singularities do not appear in the
perturbative calculations, but should also find solutions that are
insensitive to soft radiation in the event. \ As illustrated in
Fig.~\ref{IRSafe}, algorithms that look for jets only around towers
that exhibit some minimum energy activity, called seed towers or
just seeds, can be quite sensitive to soft radiation. The
experimental cone algorithms employed in previous runs have
such seeds.

\item  {\em Collinear safety:} The algorithm should not only be collinear
safe, in the sense that collinear singularities do not appear in
the perturbative calculations, but should also find jets that are
insensitive to any collinear radiation in the event.

A) Seed-based algorithms will in general break collinear safety
until the jets are of sufficiently large $E_{T}$ that splitting of
the seed energy between towers does not affect jet finding (See
Fig.~\ref{COsafe}). This was found to be the case for jets above 20
GeV in the D\O\ data, where jets were found with 100\% efficiency
using a seed tower threshold of 1.0~GeV~\cite{D0note3324}.
The collinear dependence introduced via the
seed threshold is removed when the jets have sufficient $E_{T}$ to
be reconstructed with 100\% efficiency.

B) Another possible collinear problem can arise if the algorithm is
sensitive to the $E_{T}$ ordering of particles.  An example would
be an algorithm where a) seeds are treated in order of decreasing $E_T$
and b) a seed is removed from the seed list when it is within a jet
found using a seed that is higher on the list.  For such an algorithm
consider the configuration illustrated in
Fig.~\ref{COsafe2}.  The difference between the two
situations is that the central (hardest) parton splits into two
almost collinear partons. The separation
between the two most distant partons is more than $R$ but less
than $2 R$.  Thus all of the partons can fall within a single cone
of radius $R$ around the central parton(s).  However, if the partons
are treated as seeds and analyzed with the candidate algorithm
suggested above,
different jets will be identified in the two situations.  On the
left, where the single central parton has the largest $E_T$, a single jet
containing all three partons will be found.  In the situation on the right,
the splitting of the central parton leaves the right-most parton with
the largest $E_T$.  Hence this seed is looked at first and a jet may be found
containing only the right-most and two central partons.  The left-most parton
is a jet by itself.  In this case the jet number changes depending on the
presence or absence of a collinear splitting.  This signals an incomplete
cancellation of the divergences in the real and virtual contributions
to this configuration and renders the algorithm collinear unsafe.
While the algorithm described here is admittedly an extreme case,
it is not so different from some schemes used in Run I.  Clearly
this problem should be avoided by making the selection or ordering of
seeds and jet cones independent of the $E_{T}$ of individual
particles.

\item  {\em Invariance under boosts:} The algorithm should find the
same solutions independent of boosts in the longitudinal direction.
\ This is particularly important for $p\overline{p}$ collisions
where the center-of-mass of the individual parton-parton collisions
is typically boosted with respect to the $p\overline{p}$
center-of-mass. This point was emphasized in conversations with the
Jet Definition Group Les Houches~\cite{LESH}.\footnote{The Les Houches group
discussed jet algorithms for both the Tevatron and LHC, and they
sharpened their algorithm requirements by also requiring boundary
stability (the kinematic boundary for the one jet inclusive jet
cross section should be at the same place, $E_{T}=\mbox{$\sqrt{\rm
s}$}/2$, independent of the number of final state particles),
suitability for soft gluon summations of the theory, and simplicity
and elegance.}

\item  {\em Boundary Stability:} It is desirable
that the kinematic variables used to describe the jets exhibit
kinematic boundaries that are insensitive to the details of the
final state. For example, the scalar $E_{T}$ variable,
explained in more detail in the next section, has a boundary that
is sensitive to the number of particles present and their relative
angle ({\it i.e}., the boundary is sensitive to the mass of the
jet). The bound$\ E_{T}^{max}=\sqrt{s}/2$ applies only for
collinear particles and massless jets. \ In the case of massive
jets the boundary for $E_{T}$ is larger than $\sqrt{s}/2$.
Boundary stability is essential in order to perform soft gluon summations.

\item  {\em Order Independence:} The algorithm should find the
same jets at parton, particle, and detector level. \ This feature
is clearly desirable from the standpoint of both theory and
experiment.

\item  {\em Straightforward Implementation:} The algorithm should
be straightforward to implement in perturbative calculations.
\end{enumerate}

\subsection{Experimental Attributes of the Ideal \\ Algorithm}

Once jets enter a detector, the effects of particle showering, detector
response, noise, and energy from additional hard scatterings from the same
beam crossing will subtly affect the performance of even the most ideal
algorithm. It is the goal of the experimental groups to correct for such
effects in each jet analysis. Ideally the algorithm employed should not
cause the corrections to be excessively large. From an ``experimental
standpoint'' we add the following criteria for a desirable jet algorithm:

\begin{enumerate}
\item  {\em Detector independence:} The performance of the algorithm
should be as
independent as possible of the detector that provides the data. For example,
the algorithm should not be strongly dependent on detector segmentation,
energy response, or resolution.

\item  {\em Minimization of resolution smearing and angle biases:}
The algorithm should not amplify the inevitable effects of
resolution smearing and angle biases.

\item  {\em Stability with luminosity:} Jet finding should not be strongly
affected by multiple hard scatterings at high beam luminosities.
For example, jets should not grow to excessively large sizes due to
additional interactions.  Furthermore the jet angular and energy
resolutions should not depend strongly on luminosity.

\item  {\em Efficient use of computer resources:} The jet algorithm should
provide jet identification with a minimum of computer time.  However,
changes in the algorithm intended to minimize the necessary computer
resources, {\it e.g.},
the use of seeds and preclustering, can lead to problems in the
comparison with theory.  In general, it is better to invest in more
computer resources instead of distorting the definition of the algorithm.

\item  {\em Maximal reconstruction efficiency:} The jet algorithm should
efficiently identify all physically interesting jets ({\it i.e}., jets
arising from the energetic partons described by perturbative QCD).

\item  {\em Ease of calibration:} The algorithm should not present
obstacles to
the reliable calibration of the final kinematic properties of the jet.

\item  {\em Ease of use:} The algorithm should be straightforward to
implement
with typical experimental detectors and data.

\item {\em Fully specified:} Finally, the algorithm must be
{\bf fully specified}. \ This includes specifications for
clustering, energy and angle definition, and all details of jet
splitting and merging.

\end{enumerate}

These experimental requirements are primarily a matter of optimization
under real-life conditions and will, in general, exhibit complicated
sensitivities to running conditions.
They have a strong bearing on the ease with which
quality physics measurements are achieved.  Many of the details
necessary to fully implement the jet algorithms have neither been
standardized nor
widely discussed and this has sometimes led to misunderstandings
and confusion. The remainder of this chapter describes the cone and
\KT algorithms discussed and recommended by the QCD at
Run II Jets Group.

\section{Cone Jet Algorithms}

\subsection{Introduction}

This section should serve as a guide for the definition of common
cone jet algorithms for the Tevatron and possibly future
experiments.  Section \ref{ss:definitions} reviews the features of
previously employed cone algorithms. Section~\ref{ss_Seedless}
describes a seedless cone algorithm. Section~\ref{ss_Seed} gives a
description of seed-based cone algorithms and discusses the need
for adding midpoints between seeds as alternate starting points for
clustering. Finally, in Section~\ref{ss_Prop}, we offer a detailed
proposal for a common cone jet algorithm in Run~II analyses.

\subsection{Review of Cone Algorithms}

\label{ss:definitions}

Cone algorithms form jets by associating together particles whose
trajectories ({\it i.e}., towers whose centers) lie within a
circle of specific radius $R$ in $\eta \times \phi $ space. This
2-dimensional space is natural in $p\overline{p}$ collisions where
the dynamics are spread out in the longitudinal direction. \
Starting with a trial geometric center (or axis) for a cone in
$\eta \times \phi $ space, the energy-weighted centroid
is calculated including contributions from all particles within
the cone. This new point in $\eta \times \phi $ is then used as the
center for a new trial cone. \ As this calculation is iterated the
cone center ``flows'' until a ``stable'' solution is found, {\it
i.e}., until the centroid of the energy depositions within the cone
is aligned with the geometric axis of the cone. This leads us to
our initial cone algorithm based on the Snowmass~
scheme~\cite{SNOWMASS} of scalar $E_{T}$-weighted centers. \ The
particles are specified by massless 4-vectors $(E^{i}=|{\bf
p}^{i}|,{{\bf p}^{i}})$ with angles $\left( \phi
^{i},\theta
^{i},\eta ^{i}=-\ln \left( \tan (\theta ^{i}/2)\right) \right) $ given by
the direction from the interaction point with unit vector
$\hat{{\bf p}}^{i}={\bf p}^{i}/E^{i}$. \ The scalar $E_{T}$ for
each particle is $E_{T}^{i}=E^{i}\sin (\theta ^{i})$. \ For a
specified geometric center for the cone $\left( \eta ^{C},\phi
^{C}\right) $ the particles $i$ within the cone satisfy
\begin{equation}
i\subset C\quad  : \quad\sqrt{\left( \eta ^{i}-\eta ^{C}\right)
^{2}+\left( \phi ^{i}-\phi ^{C}\right) ^{2}}\leq R.
\end{equation}
In the Snowmass algorithm a ``stable'' cone
(and potential jet) satisfies the constraints
\begin{equation}
\eta ^{C}=\frac{\sum_{i\subset C}E_{T}^{i}\eta ^{i}}{E_{T}^{C}},\quad \phi
^{C}=\frac{\sum_{i\subset C}E_{T}^{i}\phi ^{i}}{E_{T}^{C}}  \label{centroid}
\end{equation}
({\it i.e}., the geometric center of the previous equation is identical to
the $E_{T}$-weighted centroid) with
\begin{equation}
E_{T}^{C}=\sum_{i\subset C}E_{T}^{i}\;.
\end{equation}
Naively we can simply identify these stable cones, and the
particles inside, as jets, $J=C$. \ (We will return to the
practical issues of the impact of seeds and of cone overlap below.)

To complete the jet finding process we require a recombination
scheme. Various choices for this recombination step include:

\begin{enumerate}
\item  {\em Original Snowmass scheme:} Use the stable cone variables:
\begin{eqnarray}
E_{T}^{J} &=&\sum_{i\subset J=C}E_{T}^{i}=E_{T}^{C} \label{et} \;,\\
\eta ^{J} &=&\frac{1}{E_{T}^{J}}\sum_{i\subset J=C}E_{T}^{i}\eta ^{i}\;,
\label{eta1} \\
\phi ^{J} &=&\frac{1}{E_{T}^{J}}\sum_{i\subset J=C}E_{T}^{i}\phi ^{i}\;.
\label{phi1}
\end{eqnarray}

\item  {\em Modified Run I recombination schemes:}
After identification of the jet as the contents of the stable cone,
construct more 4-vector-like variables:

\begin{eqnarray}
E_{x}^{i} &=&E_{T}^{i}\cdot \cos (\phi ^{i})\;, \\
E_{y}^{i} &=&E_{T}^{i}\cdot \sin (\phi ^{i})\;, \\
E_{z}^{i} &=&E^{i}\cdot \cos (\theta ^{i})\;, \\
E_{x,y,z}^{J} &=&\sum_{i\subset J=C}E_{x,y,z}^{i}\;,  \\
\theta ^{J} &=&\tan ^{-1}(\frac{\sqrt{(E_{x}^{J})^{2}+(E_{y}^{J})^{2}}}{
E_{z}^{J}})\;.
\end{eqnarray}

A) In Run I, D\O \ used the scalar $E_{T}^{J}$ sum as defined in Eq.~\ref{et}
but used the following definitions
for $\eta ^{J}$ and $\phi ^{J}$:
\begin{eqnarray}
\eta ^{J} &=&-\ln \left( \tan (\frac{\theta ^{J}}{2})\right)\;,\label{eta} \\
\phi ^{J} &=&\tan ^{-1}(\frac{E_{y}^{J}}{E_{x}^{J}})\;.\label{phi}
\end{eqnarray}

B) In Run I, CDF used the angular definitions in Eqs.~\ref{eta}--\ref{phi}
and also replaced the Snowmass scheme $E_{T}^{J}$ with:
\begin{equation}
E_{T}^{J}=E^{J}\cdot \sin (\theta ^{J}),\quad E^{J}=\sum_{i\subset J}E^{i}\;.
\end{equation}
\end{enumerate}

\noindent Note that in the Snowmass scheme the designation of the
centroid quantities $\eta ^{J}$ and $\phi ^{J}$ of Eqs.~\ref{eta1}
and \ref{phi1} as a pseudorapidity and an azimuthal
angle is purely convention. These quantities only approximate the
true kinematic properties of the massive cluster that is the jet.
They are, however, approximately equal to the ``real'' quantities,
becoming exact in the limit of small jet mass ($M^{J}<<E_{T}$).
Further these quantities transform simply under longitudinal boosts
({\it i.e.}, $\eta ^{J}$ boosts additively while $\phi ^{J}$ is
invariant) guaranteeing that the jet structure determined with the
Snowmass algorithm is boost invariant.  It is also worthwhile noting that
the Snowmass $\eta ^{J}$ is a better estimator of the ``true'' jet
rapidity ($y^J$) defined below than the ``true'' jet pseudorapidity
defined in Eq.~\ref{eta}.  The latter quantity does not boost
additively (for $M^{J}>0$) and is not a good variable for systematic
studies.

While the scalar sum $E_{T}$
is invariant under longitudinal boosts, it is not a true energy
variable. This feature leads to difficulty in resummation
calculations: the kinematic boundary of the jet $E_{T}$ shifts away
from $\sqrt{s}/2$ appropriate for two parton kinematics when
additional final state partons are included and the jet acquires a
nonzero mass. On the other hand the Snowmass variables have the
attractive feature of simplicity, involving only arithmetic rather
than transcendental relationships. \ An alternate choice, which we
recommend here, is to use full 4-vector variables for the jets.

\begin{itemize}
\item[] \hspace{-6mm}3.\hspace{4mm}{\em {E}--Scheme, or 4-vector recombination:}
\begin{eqnarray}
p^{J} &=&(E^{J},{\bf p}^{J})=\sum_{i\subset
J=C}(E^{i},p_{x}^{i},p_{y}^{i},p_{z}^{i})\;, \\
p_{T}^{J} &=&\sqrt{(p_{x}^{J})^{2}+(p_{y}^{J})^{2}}\;, \\
y^{J} &=&\frac{1}{2}\ln \frac{E^{J}+p_{z}^{J}}{E^{J}-p_{z}^{J}}\;,\quad
\phi ^{J}=\tan ^{-1}\frac{p_{y}^{J}}{p_{x}^{J}}\;.
\end{eqnarray}
\end{itemize}

\noindent Note that in this scheme one does {\it not} use the scalar $E_T$
variable.  The 4-vector variables defined above manifestly display
the desired Lorentz properties. Phase space boundaries will exhibit
the required stability necessary for all-order resummations. \
While the structure of analytic fixed order perturbative
calculations is simpler with the Snowmass variables, NLO cross
section calculations are now also possible with Monte Carlo
programs~\cite{nlo3jet,ohnow,MEPJET,disMC}. Such programs are fully
flexible with respect to the choice of variables and the 4-vector
variables pose no practical problems. It is also important to
recall that, at least at low orders in perturbation theory, it is
not possible for energy to be conserved in detail in going from the
parton level to the hadron level. At the parton level the jet will
almost surely be a cluster of partons with non-zero color charge.
At the hadron level the cluster will be composed of color-singlet
hadrons. The transition between the two levels necessarily involves
the addition (or subtraction) of at least one colored parton
carrying some amount (presumably small) of energy.

One can also employ these true 4-vector variables,
rather than the $E_{T}$-weighted centroid, in the jet algorithm to
find stable cones.\ While this choice will complicate the
analysis, replacing simple arithmetic relationships with
transcendental relationships, the group recommends that this
possibility be investigated.  The goal is to have a uniform set
of kinematic variables with appropriate Lorentz properties
throughout the jet analysis.

At this point it might seem that a simple and straightforward jet
definition would arise from just the choice of a cone size and a
recombination scheme. The algorithm would then be used to scan the
detector and simply find all stable cones. \ In practice, this
naive algorithm was found to be incomplete. \ To keep the time for
data analysis within reasonable bounds the concept of the seed was
introduced. \ Instead of looking ``everywhere'' for stable cones,
the iteration process started only at the centers of seed towers
that passed a minimum energy cut (how could a jet not have sizeable
energy deposited near its center?).  Additionally, in {\mbox Run~I}
both CDF and \D0 reduced the number of seed towers used as starting
points by consolidating adjacent seed towers into single starting
points. (The actual clustering was always performed on calorimeter
towers.) \ These types of procedures, however, create the problems
illustrated in Figs.~\ref{IRSafe}, \ref{COsafe} and \ref{COsafe2},
introducing sensitivity to soft emissions and the possibility of
collinear sensitivity.

The naive Snowmass algorithm also does not address the question of
treating overlapping stable cones. \ It is quite common for two
stable cones to share some subset (but not all) of their particles.
\ While not all particles in the final state need to be assigned to
a jet, particles should not be assigned to more than one jet. \
Hence there must be a step between the stable cone stage and the
final jet stage where either the overlapping cones are merged (when
there is a good deal of overlap) or the shared particles are split
between the cones. \ Typically cones whose shared energy is larger
than a fixed fraction ({\it e.g}., $f = 50\%$) of the energy in
the lower energy cone are merged. \ For the cases with shared
energy below this cut, the shared particles are typically assigned
to the cone that is closer in $\eta \times \phi $ space. \ As
suggested earlier, the detailed properties of the final jets will
depend on the merge/split step and it is essential that these
details be spelled out in the algorithm. \ We provide examples
in the following sections.

\subsection{Cone Jets without Seeds}

\label{ss_Seedless}

Since many of the issues outlined in the previous section arise
from the use of seed towers to define the starting point in the
search for stable cones, it is worthwhile to consider the
possibility of a seedless cone algorithm. \ A seedless algorithm
is infrared insensitive.  It searches the entire
detector and finds all stable cones (or
proto-jets\footnote{ At the clustering stage we refer to stable
cones as proto-jets. These may be promoted to jets after surviving
the splitting and merging stage.}), even if these cones do not have a
seed tower at their center. Collinear sensitivity is also removed,
because the structure of the energy depositions within the cone is
unimportant. \ In this section we present a preliminary study
of such an algorithm.

\subsubsection{Seedless Jet Clustering}

\label{sec:seedless}

\begin{figure}[tbp]
\begin{center}
\epsfig{figure=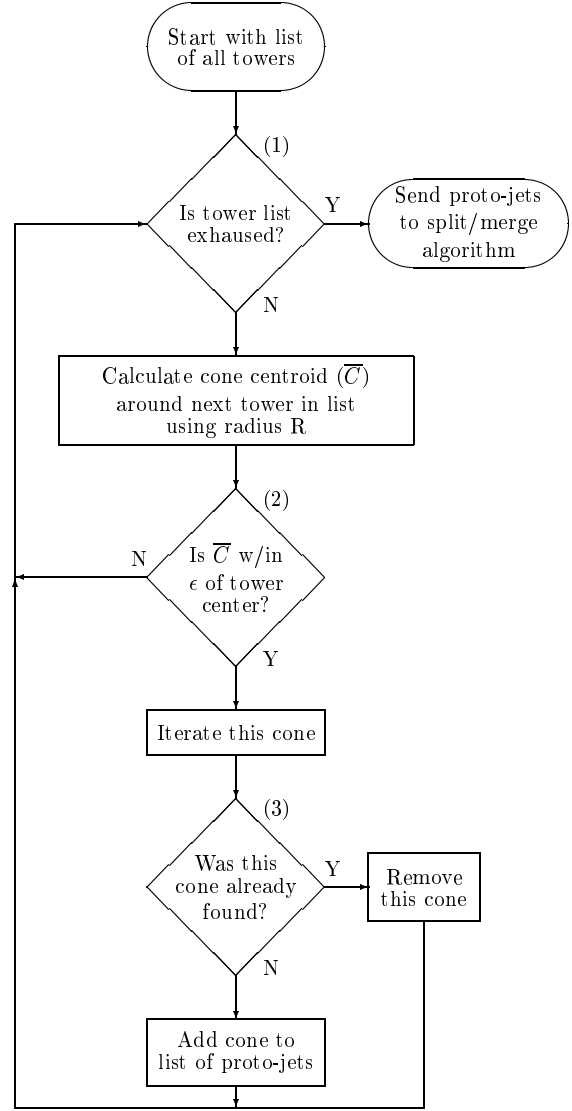,width = 75mm}
\end{center}
\caption{A seedless clustering algorithm.}
\label{seedless_1}
\end{figure}

We give an example of a seedless algorithm in the flowchart in
Fig.~\ref {seedless_1}. \ The basic idea~\cite{seedsoper}
follows from the concept
of ``flowing'' cone centers mentioned earlier. \ The location of a
stable cone will act as an attractor towards which cones will flow
during the iteration process. \ If the process starts close to such a
stable center, the flow steps will be small. \ Starting points further
from a stable center will exhibit larger flow steps towards the stable
center during the
iteration. Starting points outside of the region of attraction will
again exhibit small flow steps. \ The method starts by looping
through {\it all}
detector towers\footnote{While the algorithm may be run on
individual detector cells, we do not believe that cell-level
clustering is within the CPU means of current experiments for the
largest expected data samples.} in some appropriate fiducial volume.
For each tower $k$, with center
$\overrightarrow{k}=$ $\left( \eta ^{k},\phi ^{k}\right) $, we
define a cone of size $R$ centered on the tower
\begin{eqnarray}
\overrightarrow{C^{k}} &=&\left( \eta ^{C^{k}}=\eta^{k},\phi ^{C^{k}}=\phi^{k}\right) ,\nonumber \\
i &\subset &C^{k}\; : \; \sqrt{\left( \eta ^{i}-\eta^{C^{k}}\right) ^{2}+\left(
\phi ^{i}-\phi^{C^{k}}\right) ^{2}}\leq R .\label{flow}
\end{eqnarray}
For each cone we evaluate the $E_{T}$-weighted centroid
\begin{eqnarray}
\overrightarrow{\bar{C}^{k}}&=&\left( \bar{\eta}^{C^{k}},\bar{\phi}^{C^{k}}\right) \;, \label{centroid2} \\
\bar{\eta}^{C^{k}} &=&\frac{\sum_{i\subset C^{k}}E_{T}^{i}\eta ^{i}}{E_{T}^{C^{k}}},\; \bar{\phi}
^{C^{k}}=\frac{\sum_{i\subset C^{k}}E_{T}^{i}\phi ^{i}}{E_{T}^{C^{k}}}\;,
\\
E_{T}^{C^{k}} &=&\sum_{i\subset C^{k}}E_{T}^{i}\;.
\end{eqnarray}
Note that, in general, the centroid $\overrightarrow{\bar{C}^{k}}$
is not identical to
the geometric center $\overrightarrow{C^{k}}$ and the cone is not stable.
While this first step is resource intensive,
we simplify the subsequent analysis with the next step. \
If the calculated centroid of the cone lies outside of the initial
tower,
further processing of that cone is skipped and the cone is
discarded. The specific exclusion distance used in this cut is a
somewhat arbitrary parameter and could be adjusted to maximize jet
finding efficiency and minimize the CPU demand of the algorithm.
All cones that yield a centroid within the original tower become
preproto-jets. For these cones the process of calculating a new
centroid about the previous centroid is iterated
and the cones are allowed to ``flow'' away from the original towers.
This iteration continues until either a stable cone center is found
or the centroid migrates
out of the fiducial volume. The surviving stable cones constitute
the list of proto-jets. \ Note that the tower content of a cone
will vary as its center moves within the area of a single tower. \
For a cone of radius $R$ and tower dimension $\Delta $ (in either
$\eta $ or $\phi $) the minimum change in the cone center location
for which the tower content in the cone
changes by at least one tower is characterized by $\Delta
^{2}/2R $.  This distance is of order $0.007$ for $\Delta=0.1$ and $R=0.7$
({\it i.e}., 10\% of a tower width if the diameter of the cone,
$2R,$ is ten times a tower width).

An even more streamlined option would be to keep only those cones
that yield a stable cone center without leaving the original tower.
Since a trial cone is originally placed at the center of every
tower, the only distinct stable cone centers missed by this (much!)
faster algorithm correspond to very limited regions of attraction
(less than the area of a tower). Such situations can arise in only
two cases. One possibility is that there are two (or more) stable
directions within a single tower. The second possibility is that
there is a stable direction within a tower but it is not found
starting at the tower center. While both of these scenarios arise
in analyses of realistic data, they do not constitute cause for
concern. Proto-jets with directions that are nearly collinear ({\it
i.e.}, that lie within a single tower) will have nearly the same tower
content and be merged with little impact on the final jet
properties. Isolated stable directions with very small regions of
attraction (the second case) are most likely fluctuations in the
background energy level and not the fingerprints of real emitted
partons. In any case the stable cone centers not found by the
streamlined algorithm invariably correspond to low $E_{T}$
proto-jets and are well isolated from large $E_{T}$ proto-jet
directions (otherwise they would be attracted into the larger
$E_{T}$ jet). Thus the leading $E_{T}$ jets (after merging and
splitting) found by either the original seedless algorithm or the
streamlined version are nearly identical.

For practical use it may also be necessary to apply some minimum
$E_T$ threshold to the list of proto-jets. Ideally such a threshold
would be set near the noise level of the detector. However, a
higher setting might be warranted to reduce the sensitivity of the
algorithm to energy depositions by multiple interactions at high
luminosities (see Section \ref{conelessdata} for details of
seedless clustering at the detector level).

In general, a number of overlapping cones, where towers are shared by more
than one cone, will be found after applying the stable cone finding
procedure. As noted earlier, the treatment of proto-jets with overlapping
regions can have significant impact on the behavior of the algorithm.

\subsubsection{Splitting and Merging Specifications}

\begin{figure}[tbp]
\begin{center}
\epsfig{figure=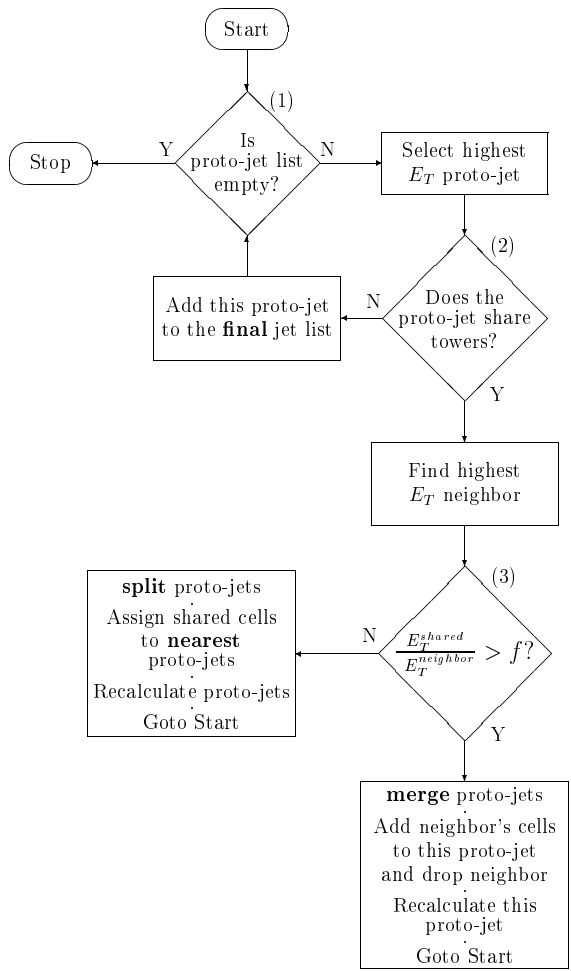,width = 75mm}
\end{center}
\caption{A fully specified splitting and merging algorithm.}
\label{split_merge}
\end{figure}

A well-defined algorithm must include a detailed prescription for
the splitting and merging of proto-jets with overlapping cones. We
provide an outline of a splitting and merging algorithm in
Fig.~\ref{split_merge}. It is important to note that the splitting
and merging process does not begin until all stable cones have been
found. Further, the suggested algorithm always works with the
highest $E_{T}$ proto-jet remaining on the list and the ordering of
the list is checked after each instance of merging or splitting. If
these conditions are not met, it is difficult to predict the
behavior of the algorithm for multiply split and/or merged jets and
similar lists of proto-jets can lead to distinctly different lists
of jets. \ This undesirable situation does not arise with the well-
ordered algorithm in Fig.~\ref{split_merge}. While there will
always be some order dependence in a splitting and merging scheme
when treating multiply overlapping jets, we recommend fixing this
order by starting with the highest $E_{T}$ proto-jet and working
down in the $E_{T}$ ordered list. In this way the action of the
algorithm is to prefer cones of maximal $E_{T}$. Note that, after a
merging or splitting event, the $E_{T}$ ordering on the list of
remaining proto-jets can change, since the survivor of merged jets
may move up while split jets may move down. Once a proto-jet shares
no towers with any of the other proto-jets, it becomes a jet and is
not impacted by the subsequent merging and splitting of the
remaining proto-jets. As noted earlier and illustrated in
Fig.~\ref{split_merge}, the decision to split or merge a pair of
overlapping proto-jets is based on the percentage of
transverse energy shared
by the lower $E_{T}$ proto-jet. Proto-jets sharing a fraction
greater than $f$ (typically $f=50\%$) will be merged; others will
be split with the shared towers individually assigned to
the proto-jet that is closest in $\eta \times \phi$ space.
This method will perform predictably even in the case of
multiply split and merged jets.  Note that there is no requirement
that the centroid of the split or merged proto-jet still
coincides precisely with its geometric center.

\subsubsection{Parton Recombination}

\label{sec:seedless.parton}

The definition of calorimeter towers, {\it i.e.}, a discretization
of $(\eta ,\phi )$ space, would be cumbersome in a theoretical
calculation, and is indeed not necessary. In a theoretical
calculation at fixed order, the maximal number of partons, $n$, is
fixed. With specified parton momenta, the only possible positions
of stable cones are then given by the partitions of the $n$ parton
momenta, {\it i.e.}, there are at most $2^{n}-1$ possible locations
of proto-jets. They are given by the positions of individual
partons, all pairs of partons, all combinations of three partons,
{\it etc}. In a perturbative calculation, {\it e.g}. via a NLO Monte
Carlo program, the proto-jet selection of the seedless algorithm
can then be defined as follows:

\begin{enumerate}
\item  Make a list of centroids for all possible parton multiplets.
These are derived from the coordinates of all parton
momenta $p_{i}$, of all pairs of parton momenta $p_{i}+p_{j}$, of
all triplets of parton momenta $ p_{i}+p_{j}+p_{k}$, {\em etc}. For each
centroid record which set of partons defines it.

\item
Select the next centroid on the list as the center of a trial
cone of radius $R$.\newline
Go to the split/merge stage if the list of cone centers is exhausted.

\item  Check which partons are inside the trial cone.

\item  If the parton list of the centroid and that of
the trial cone disagree, discard the trial cone and go to (2). If
the lists agree, add the set of partons inside the trial cone as a
new entry to the list of proto-jets.
\end{enumerate}

\noindent As before, different proto-jets may share partons, {\it i.e}. they
may overlap. The required split/merge step is then identical to the
calorimeter-level steps (Fig.~\ref{split_merge}), with towers replaced by
partons as elements of proto-jets.

In the case of analytic evaluations of the NLO perturbative jet
cross section~\cite{EKS} the integrations over the multi-parton
phase space are divided into various disjoint contributions. For a
jet of fixed $E_{T}^{J}$, $\eta ^{J}$ and $\phi ^{J}$ we have only
the cases where a) one parton is in the jet direction with the jet
$E_{T}$, and the other partons are excluded from nearby directions
where they could fit in a jet cone with the first parton, or b) two
partons fit in a single cone with their centroid properties
constrained to be the jet values. The questions of overlap,
splitting and merging never arise at this order for $R<\pi /3$.

\subsubsection{Tests of a Seedless Algorithm}

\label{conelessdata}

\begin{figure}[t]
\epsfig{figure=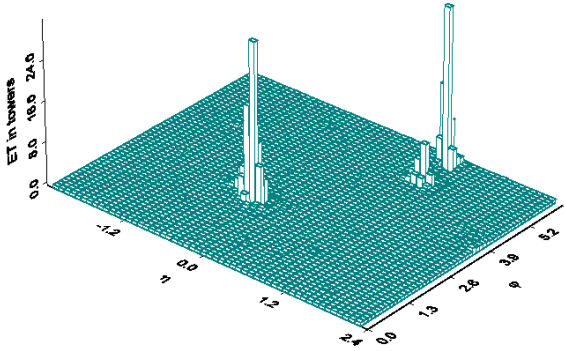,width=75mm}
\caption{Calorimeter tower $E_T$ lego plot for a simulated large-$E_T$ jet
event in the D\O\ Calorimeter.}
\label{fig:PyTower1605}
\end{figure}

\begin{figure}[ht]
\epsfig{figure=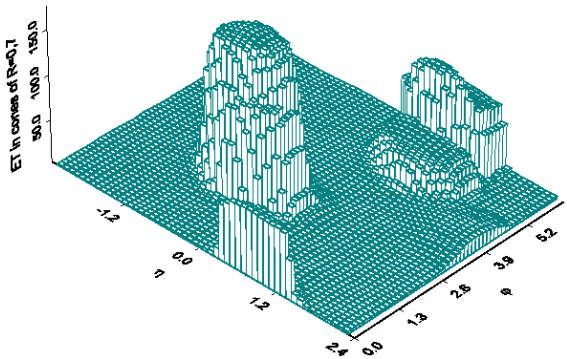,width=75mm}
\caption{$E_T$ in cones centered on each calorimeter tower (in $|\protect\eta
^{tower}|<2.4$) for the simulated large-$E_T$ jet event of Fig.~\ref
{fig:PyTower1605}.}
\label{fig:PyCone1605}
\end{figure}

\begin{figure}[tbp]
\epsfig{figure=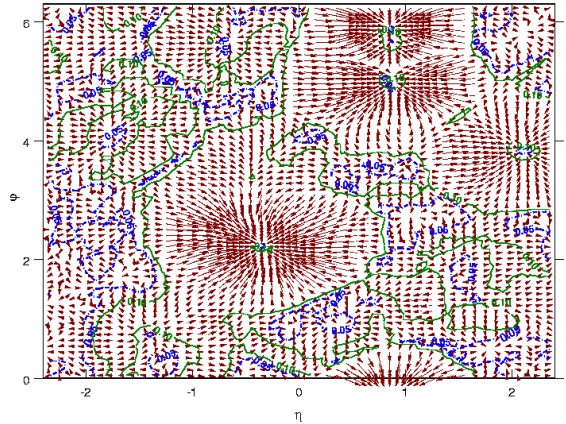,width=75mm}
\caption{Energy flow for the cones in the large-$E_T$ jet event of Figs.~\ref
{fig:PyTower1605}--\ref{fig:PyCone1605}. The contours bound flow
regions with vector magnitude $< 0.1$ (solid contours) and
$<0.05$ (dashed contours) in $\protect\eta\times\protect\phi$.}
\label{fig:PyFlow1605}
\end{figure}

In this section we offer some insight into the performance of the
seedless cone algorithm applied to a detector. We begin by
examining a simulated large-$E_{T}$ jet event in the D\O\ detector
(Fig.~\ref{fig:PyTower1605}). The event was chosen from a sample
generated with {\sc pythia}~\cite{Pythia} using a 160 GeV minimum
$E_{T}$ cut at the parton-level generator. After hadronization, the
events were processed through a full simulation of the D\O\
detector. The towers in the central region ($-3.2<\eta <3.2$) are
$0.1\times 0.1$ in size. Fig.~\ref {fig:PyTower1605} shows the
distribution of calorimeter tower $E_{T}$'s for the event in the
central fiducial volume ($-2.4<\eta <2.4$) where cones of $R=0.7$
can be fully contained in the central region. Three jets clearly
dominate the display (along with a less distinctive feature at the
large $\eta $ boundary near $\phi =4$). Fig.~\ref{fig:PyCone1605}
shows the $E_{T} $ contained in a cone of radius 0.7 centered at
each calorimeter tower, displaying the same structure for the event
in a slightly different language. We can make this picture even
more clear by appealing to the ``flow imagery'' of
Section~\ref{sec:seedless}. We define a flow vector as the
2-dimensional vector difference between the calculated centroid for
a cone centered on a tower and the geometric center of the tower
($\overrightarrow{\bar{C}^{k}}-\overrightarrow{C^{k}}$ in
Eqs.~\ref{flow} and \ref{centroid2}). This vector vanishes for a
stable cone. This flow vector is plotted in the corresponding range
of $\eta
\times \phi $ in Fig.~\ref{fig:PyFlow1605} for the same {\sc pythia}
generated event.

The flow vector clearly points to the four potential jets noted above.
Cones that are in the neighborhood of a potential jet exhibit flow vectors of
large magnitude pointing towards the jet center. This
magnitude will generally be sufficient to cause the cone to fail
the second test in Fig.~\ref{seedless_1}, thus preventing further
iteration of the cone to define a proto-jet. The contours of
Fig.~\ref{fig:PyFlow1605} bound regions of flow with magnitude
$<0.1$ (solid contours) and $<0.05$ (dashed contours) in $\eta
\times\phi $, within which we expect to find the final jets. It is
important to note the size of the detector regions with small flow
magnitude. Regions with sufficiently small flow will pass test~(2)
in the clustering stage and allow the cone to undergo additional
iterations. \ This ultimately increases processing time for
clustering and complexity in splitting and merging (due to the
production of many additional proto-jets). The flow magnitude cut
has a natural size on the order of the detector tower size. \ For
the D\O\ detector, with a typical towers size of $\eta \times \phi
= 0.1 \times 0.1$, the cut would be between the two contours shown above.
A too small magnitude cut will cause inefficiencies in jet finding;
too large a cut will cause iterations on cones over the whole
detector volume.

It is clear from Figs.~\ref{fig:PyTower1605}--\ref{fig:PyFlow1605}
that the region of interest around the jets is much smaller than
the area contained within the contours of ``stable'' cones. \ There
are broad ``plains'' of low energy deposition where the flow vector
is of small magnitude, but also of rapidly varying direction. \
Stable cones are found in these regions. \ But these presumably
arise simply from local fluctuations yielding local extrema and are
not expected to correspond to the fingerprints of underlying
(energetic) partons. \ There are at least two, possibly parallel
paths to follow in order to reduce the impact of these regions on
the analysis, in terms of both required resources and final
results.

As already noted, we can further streamline the analysis by
applying the cut on the flow vector at each step in the iteration.
\ Thus we keep only those cones that do not ``flow'' outside of
their original tower before a stable center is reached. \ Such an
algorithm converges rapidly to the stable cones pointed to by the
largest magnitude flow vectors in Fig.~\ref{fig:PyFlow1605} and
efficiently eliminates most of the cones in the ``plains''. We do
lose the stable cones that a full iteration, allowing any amount of
flow, finds in the flat regions of the previous figures. However,
as already emphasized, these cones do not correspond to the physics
we wish to study with jet analyses. \ With a large savings in
analysis time the streamlined algorithm finds the same leading jet
properties ({\it e.g.}, $E_T$ and $\eta^J$) as the more complete
algorithm to a fraction of a percent. The final jets contain
typically 120 to 160 towers.  The differences between the leading
jets found with the two algorithms arise from differences in tower
content of just 1 or 2 towers (at the cone boundary).

One can also reduce the effort and the final event complexity by applying a
minimum $E_{T}$ cut on the cones at the proto-jet stage. An obvious choice
for this minimum $E_{T}$ cut would be to place it above the level of
detector noise. As alluded to in Section~\ref{sec:seedless}, a practical cut
might be placed slightly higher to reduce sensitivity to varying event
pileup with changes in beam luminosity. Unfortunately, this places a rather
arbitrary threshold into the algorithm from the standpoint of theoretical
calculations, {\it i.e}. what is the `noise' level at NLO? Additionally,
such cuts will in practice be applied before final jet scale corrections.
How does X\thinspace GeV uncorrected in the experiment compare to
X\thinspace GeV at generator level? Such experiment specific considerations
clearly are out of the realm of event generator design! A possible
improvement would be to set a minimum cone $E_{T}$ threshold equal to some
fraction of the scalar $E_{T}$ in the event. In this way such effects will
tend to partially cancel between generators and experiments, better relating
the cut between the two levels.

\begin{figure}[tbp]
\epsfig{figure=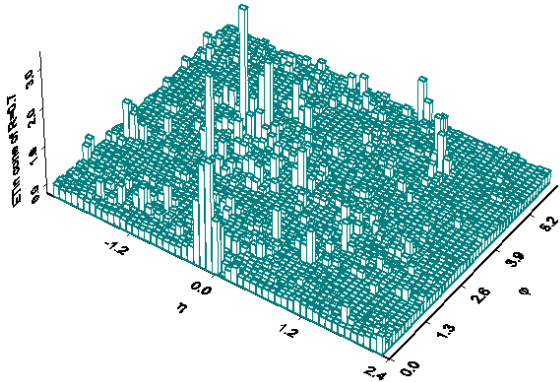,width=75mm}
\caption{A sample event from data. Tower $E_{T}$ lego plot for an event
passing the D\O\ $W\rightarrow jets$ trigger.}
\label{fig:Wjj1tower}
\end{figure}

\begin{figure}[tbp]
\epsfig{figure=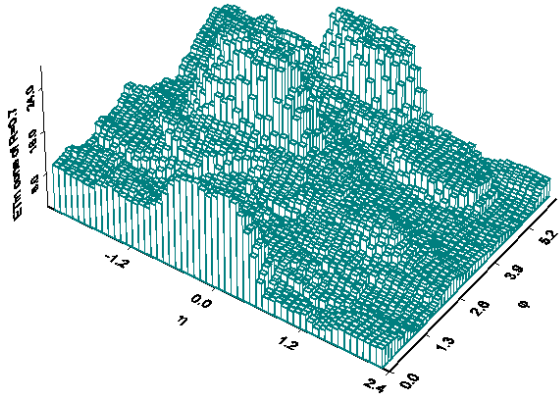,width=75mm}
\caption{$E_{T}$ in cones centered on each calorimeter tower (in $|\protect
\eta ^{T}|<2.4$) for the $W\rightarrow jets$ sample event of Fig.~\ref
{fig:Wjj1tower}.}
\label{fig:Wjj1cone}
\end{figure}

\begin{figure}[tbp]
\epsfig{figure=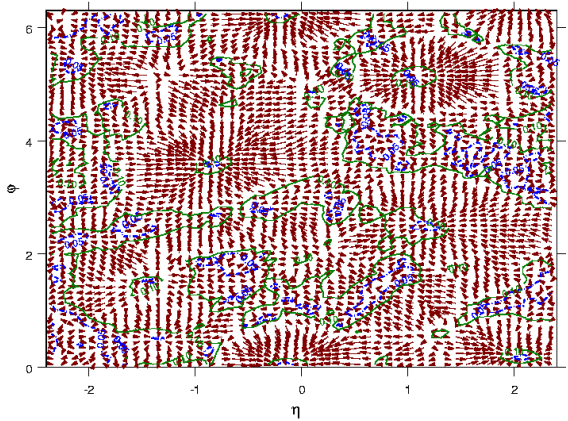,width=75mm}
\caption{Energy flow for the cones in the $W\rightarrow jets$
event of Figs.~\ref{fig:Wjj1tower}--\ref{fig:Wjj1cone}. The
contours bound flow regions with vector magnitude $<0.1$ (solid
contours) and $<0.05$ (dashed contours) in $\protect\eta \times
\protect\phi $.}
\label{fig:Wjj1flow}
\end{figure}

We next look at an example of the seedless algorithm tested on
actual calorimeter data. Fig.~\ref{fig:Wjj1tower} shows the tower
$E_{T}$ lego plot for a D\O\ event passing a $W\rightarrow jets$
trigger. The trigger required at least two central jets with
$E_{T}>15$ GeV. These data were taken at high luminosity with an
average of $\sim$2.8 interactions per beam crossing. \ The two
leading jets that pass the cut are reasonably obvious (along with,
perhaps, two other subleading jets) but overall this event is
clearly noisier (more realistic) than the {\sc pythia} generated
event. \ This point is illustrated also in Figs.~\ref{fig:Wjj1cone}
and~\ref{fig:Wjj1flow}, which show the cone energy and flow vectors
for this event, analogous to Figs.~\ref{fig:PyCone1605}
and~\ref{fig:PyFlow1605}. \ In this case the baseline energy
subtraction for calorimeter cell energies in the data leads to
towers with (small) negative energy deposition.

The increased level of noise and the possibility of negative tower
energy results in two new issues for the jet analysis that were not
observed in the analysis of the Monte Carlo data. \ The negative
energy cells allow true stability with respect to the iteration
process to be replaced by limit cycles. \ Iteration leads not only
to cone center locations for which
$\overrightarrow{\bar{C}^{j}}-\overrightarrow{C^{j}}=0$ but also, for
example, to doublets of locations for which
$\overrightarrow{\bar{C}^{{1}}}=
\overrightarrow{C^{{2}}}$ and
$\overrightarrow{\bar{C}^{{2}}}=\overrightarrow{C^{{1}}}$,
or $\overrightarrow{\bar{C}^{{1}}}-
\overrightarrow{C^{{1}}}
=-(\overrightarrow{\bar{C}^{{2}}}-\overrightarrow{C^{{2}}})$.
Thus continued iteration simply carries the cone center back and
forth between location 1 ($\overrightarrow{C^{{1}}}$) and location 2
($\overrightarrow{C^{{2}}}$). (More complex multiplets of locations
with sets of 3, or even 6, 2-dimensional flow vectors summing to 0
are also observed.) \ The good news is that these clusters of cone
centers are typically close by each other and yield essentially the
same final jets, after merging, independent of where in the limit
cycle the iteration process is terminated. \ This is guaranteed to
be true for the streamlined algorithm where the entire cycle must
occur within a single tower. \ (The $\left( \eta \times \phi
\right) $ distance between two members of such a limiting cycle
driven by a negative tower energy of magnitude $E_{N}$ {is
approximately $R\cdot E_{N}/E_{C}$, where $E_{C}$ is the total
energy in the cone. \ This can be as small as the minimum distance
for a change of one tower in the cone as noted above, {\it i.e}.,
7\% of a tower width.)

The noisy quality of the event leads to an even more troubling
phenomenon. \ There are so many locally stable cone centers found
in the now rapidly fluctuating ``plain'' region that the proto-jet list may
exhibit a surprisingly large number of mutually overlapping cones.
\ During the merging phase these can coalesce into jets with large (even
leading) $E_{T}$. \ This issue has historically been treated by applying a
minimum $E_{T}$ cut to the proto-jet list before merging and splitting. \
With the event studied here a cut of 8 GeV (typical of values used by D\O )
is not sufficient. \ If we keep all stable cones with $E_{T}>8$ GeV, with no
other cuts, as proto-jets, the merging process builds a leading jet by
pulling together many cones where there is clearly no real jet. \ This
problem does not arise in the streamlined algorithm where only stable cones
that stayed within their original tower are kept. \ In this case the
algorithm identifies the leading jets anticipated intuitively from the above
figures.

\subsubsection{Comments on the Seedless Clustering}

We may summarize the advantages of the seedless clustering
described above as follows:

\begin{enumerate}

\item  Avoids undesirable sensitivity to soft and collinear
radiation.

\item  Offers increased efficiency for all physically
interesting jets.

\item  Offers improved treatment of limit cycles and overlapping
cones.

\item  ``Flow cut'' method offers more efficient use of computer
resources than unrestricted seedless clustering.

\end{enumerate}

We have not investigated further improvements in the optimization
of the computational efficiency for this seedless algorithm.
However, some improvement may be gained by using the fact that
cones centered on adjacent towers are largely overlapping, thus
reducing the number of towers to sum for each new center.  Other
improvements such as region of interest (ROI) clustering may also
be explored.

\subsection{Cone Jets with Seeds}

\label{ss_Seed}

In an actual experiment the number of calorimeter towers may be very large
(order 6000 for tower sizes of $\Delta\eta\times\Delta\phi=0.1\times 0.1$
and an $\eta$ coverage of $\pm 5$ units of pseudorapidity). The above
seedless algorithm may then be expensive computationally. The question
arises whether an acceptable approximation of the seedless algorithm can be
constructed, analogous to the parton-level short cut, while considering
primarily those towers which have energy depositions above a minimal seed
threshold for finding proto-jets.

Seed-based cone algorithms offer the advantage of being comparatively
efficient in CPU time. In a typical application, detector towers are sorted
according to descending $E_T$ and only towers passing a seed cut,
\begin{equation}  \label{eq:etseed.def}
E_T^{tower} > E_T^{seed}\; ,
\end{equation}
are used as starting points for the initial jet cones. This greatly
reduces the number of cones that need to be evaluated in the
initial stage. The seed threshold $E_T^{seed}$ must be chosen
low enough so that variations of $E_T^{seed}$ lead to negligible
variations in any observable under consideration. The simple
seed-based algorithm is sensitive to both infrared or collinear effects.
However, sensitivity to the splitting of the seed $E_T$ between
multiple towers is greatly reduced for larger $E_T$ jets. As stated
above, this is true when the jet reconstruction becomes 100\%
efficient ({\it i.e.}, around 20\,GeV for jets in D\O ). For fully
efficient jet algorithms the collinear dependency is reduced to a
second-order effect, namely, the effective number of low $E_T$ proto-jets
that may engage in splitting and merging. In a typical algorithm a
minimum $E_T$ cut may also be applied to each proto-jet to prevent
excessive merging of noise and energy not associated with the hard
scattering producing the jets.

\subsubsection{Addition of Midpoints}

The seedless algorithm discussed previously can be approximated by
a seed-based algorithm with the addition of `midpoints' in the list
of starting seeds. The idea~\cite{midellis} is to duplicate the parton-level
algorithm discussed in Section~\ref{sec:seedless.parton}, but with
partons replaced by seeds. By adding a starting point for
clustering at the positions given by $p_i+p_j$, $p_i+p_j+p_k$ etc.,
the sensitivity of the algorithm to soft radiation as illustrated
in Fig.~\ref{IRSafe} is essentially removed. Since widely separated seeds
cannot be clustered to a proto-jet, it is sufficient to only
consider those midpoints where all seeds lie within a distance
\begin{equation}
\Delta R < 2.0 \cdot R_{cone}
\end{equation}
of each other.

With these changes, the resulting algorithm is quite close to those
used in Run I of the Tevatron. The main change is the inclusion of
midpoints of seeds (the $p_i+p_j$ pairs) and of centers of larger
numbers of seeds as additional seed locations for trial cones.
Two studies of the effects of adding midpoints were completed during the
workshop and are summarized below. The first checks the infrared safety of
the midpoint algorithm, also called the Improved Legacy Cone Algorithm
(ILCA), in a Monte Carlo study. The second tests the effect of adding
midpoints on the performance of the Run I D\O\ cone algorithm.

\subsubsection{Results from a Monte Carlo Study}

The request for an infrared and collinear safe jet-algorithm is
most important from the viewpoint of perturbative QCD calculations.
Unsafe algorithms simply do not permit unambiguous results, once
higher order corrections are considered~\cite{seymour,potter}. Instead
results will depend on the technical regularization procedure
adopted in a specific calculation.

The deficiencies of an unsafe algorithm will only show up at sufficiently
high order in the perturbative expansion. For example, the jet merging due
to soft gluon radiation as depicted in Fig.~\ref{IRSafe} will only become a
problem when three partons or more can be combined to a single jet. In
hadron collider processes this first happens in, for example, the
NLO corrections to three-jet production~\cite{nlo3jet}, where four-parton
final states are included
in the real emission contributions. The fourth parton is needed to provide
the necessary recoil transverse momentum to the other three partons which
may or may not form a single jet. The NLO three-jet Monte Carlo is very CPU
intensive, however, making it a cumbersome tool to investigate jet
algorithms, at present. A much faster probe is provided by the existing NLO
dijet Monte Carlos in DIS~\cite{MEPJET,disMC}.

In $ep\rightarrow ejjX$, the electron provides the necessary recoil
$p_{T}$ to the final-state partons. The real emission QCD
corrections at ${\cal O}(\alpha _{s}^{2})$ thus contain three
partons which can be close together. Their merging to a single jet,
with the concomitant loss of two-jet cross section, is a probe of the
infrared safety of the two-jet vs. one-jet classification of partonic
events. A second probe is provided by the $E_{T}$ flow inside a
jet, which has recently been modeled with up to three partons in a
single jet, for the current jets in DIS~\cite{dis.js}.

We have investigated these issues
with the {\sc mepjet} Monte Carlo~\cite{MEPJET},
which calculates dijet production in DIS at NLO. The program was run in a
kinematical range typical for HERA, $ep$ collisions at $\sqrt{s}=300$~GeV
with $Q^2>100$~GeV$^2$. Reconstructed jets were required to satisfy
\begin{equation}  \label{eq:ejj.region}
E_T > 10~{\rm GeV},\qquad -1<y<2,\qquad R_{jj}<2,
\end{equation}
\noindent where E-scheme recombination is used.
Here $R_{jj}$ is the separation of reconstructed jets in the legoplot.
Following HERA practice, we use a cone size $R=1$. Considering jets with a
maximal separation of twice the cone size enhances the statistical
significance of any splitting/merging effects in the Monte Carlo calculation.

With these settings two cone algorithms are considered to
investigate the importance of extra midpoints in the perturbative
results. The first is the seedless algorithm in its parton-level
implementation as described in Section~\ref{sec:seedless.parton},
which we here call the ``midpoint'' algorithm.  In order to test
the analog of tower threshold effects, only partons with
$E_{T,i}>E_T^{seed}$ are considered for centers of trial cones,
{\it i.e.}, trial cone centroids are the directions of these partons and
their midpoints $p_i+p_j$ and $p_i+p_j+p_k$. The second algorithm,
dubbed ``no center seed'' is identical, except that the midpoints
are left out as trial cone centers. For both algorithms, the final
splitting/merging decision is made with an $E_T$-fraction of
$f=0.75$ of the lower $E_T$ proto-jet as the dividing line.

The {\sc mepjet} program is based on the phase space slicing method, with
a parameter $s_{min}$ defining the separation between three-parton
final states on the one hand, and the virtual contributions plus
soft and collinear real emission processes (which cancel the
divergences of the virtual graphs) on the other. This dividing line
is completely arbitrary and observables should not depend on it. A
test of this requirement is shown in Fig.~\ref{ilca_smin} where the
dijet cross section within the cuts of Eq.~\ref{eq:ejj.region} is
shown as a function of $s_{min}$. Whereas the midpoint algorithm
shows $ s_{min}$-independence within the statistical errors of the
Monte Carlo (plain symbols), leaving out the midpoints between
partons leads to a pronounced decrease of the cross section as
$s_{min}$ becomes smaller.  Smaller $s_{min}$ implies that more
events are generated as explicit three-parton final states.  The
additional soft gluons act as extra seeds that tend to merge the
two jets, leaving the event
classified as a one-jet event, which does not contribute to the
plotted dijet cross section. The $s_{min}$ dependence of the ``no
center seed'' algorithm means that no perturbative prediction is
possible for this algorithm: as $s_{min}$ approaches zero, the
dijet cross section diverges logarithmically as $\log s_{min}/Q^2$.

\begin{figure}[t]
\begin{center}
\epsfig{figure=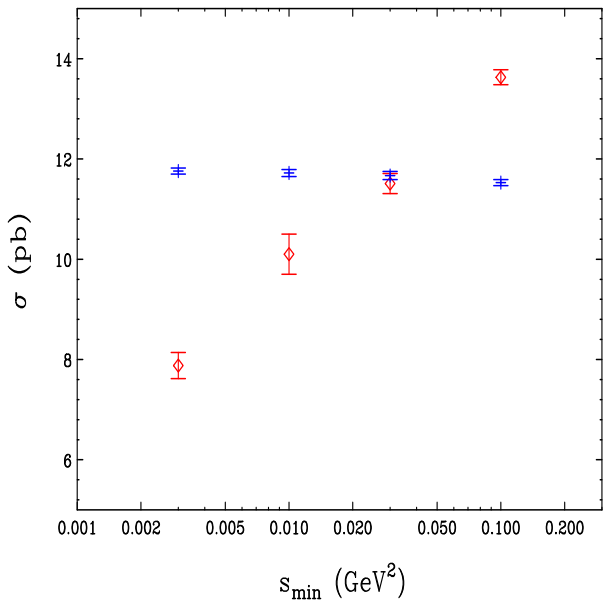,angle=90,width=80mm,height=80mm}
\end{center}
\caption{Dependence of the DIS dijet cross section on
$s_{min}$ for the ILCA algorithm with midpoints (plain
symbols) and for the ``no center seed'' algorithm (diamonds). }
\label{ilca_smin}
\end{figure}

Even when fixing $s_{min}$ to some typical soft QCD scale, like
$s_{min}=0.03\;{\rm GeV}^2$, the ``no center seed'' algorithm has
fatal defects. This is demonstrated in Fig.~\ref{ilca_eseed} where
the variation of the dijet cross section within the cuts of
Eq.~\ref{eq:ejj.region} is shown as a function of ``tower
threshold'' transverse energy $E_T^{seed}$. The midpoint algorithm
is almost independent of this threshold, as long as $E_T^{seed}$ is
less than about 10\% of the jet transverse energy. The ``no center
seed'' algorithm, on the other hand, shows a pronounced threshold
dependence, raising the specter of substantial dependence of jet
cross sections on detector thresholds, detector response to soft
particles and nonperturbative effects. These effects have been
discussed previously for three-jet events at the
Tevatron~\cite{nlo3jet,seymour}.

\begin{figure}[htb]
\begin{center}
\epsfig{figure=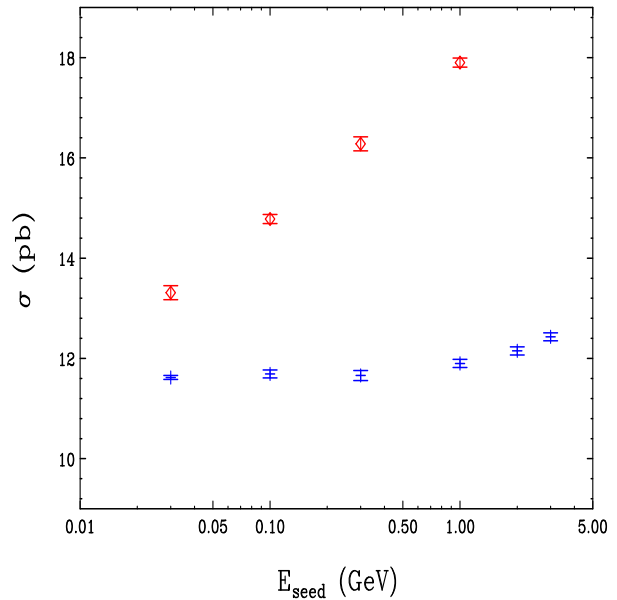,angle=90,width=80mm,height=80mm}
\end{center}
\caption{Dependence of the DIS dijet cross section on the seed threshold
$E_T^{seed}$ of Eq.~\ref{eq:etseed.def}. Results are shown for ILCA, with
midpoints (plain symbols) and for a ``no center seed'' variant (diamonds).}
\label{ilca_eseed}
\end{figure}

Discarding the ``no center seed'' algorithm we turn to internal $E_{T}$ flow
inside a single jet as another measure of the performance of jet algorithms.
The differential jet shape, $\rho (r)$, is defined as $1/\Delta r$ times the
average $E_{T}$ fraction of a jet in a narrow ring of width $\Delta r$, a
distance $r$ from the jet axis. In Fig.~\ref{jet_shapes} the differential
jet shape is shown for current jets at HERA, in the phase space region
\begin{equation}
E_{T}>14~{\rm GeV}\;,\qquad -1<\eta <2\;
\end{equation}
for DIS events with $Q^{2}>100\;{\rm GeV}^{2}$. Results are shown
for the midpoint (ILCA) and the \KT algorithm (to be described later) at
NLO ($ {\cal O}(\alpha _{s}^{2})$). The midpoint
algorithm produces wider jets than the \KT algorithm with $D=R$, as
is to be expected since two partons with a separation slightly less
than $2R$ can be clustered by the midpoint, but not the
\KT algorithm. NLO corrections are quite small for the midpoint
algorithm. We have also checked that the jet shapes in the midpoint
algorithm exhibit good scale dependence at NLO, similar to the \KT
algorithm~\cite{dis.js}.

\begin{figure}[htb]
\begin{center}
\epsfig{figure=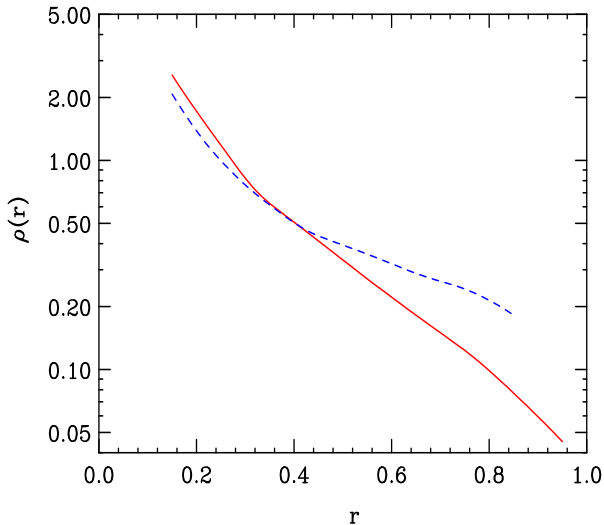,angle=90,width=80mm,height=70mm}
\end{center}
\caption{Jets shapes in ILCA (dashed line) compared to \KT (solid line).}
\label{jet_shapes}
\end{figure}

\subsubsection{Results from Data Study}

A midpoint algorithm has previously been employed by the OPAL
Collaboration~\cite{OPAL_ILCA}. We now report a study performed
using the D\O\ data. The data were acquired from a two-jet trigger
sample with an average of 2.8 interactions per beam crossing. The
goal of the data-based study was to test the sensitivity of D\O 's
Run~I cone algorithm to the addition of midpoints. To facilitate a
direct comparison of Run~II jet results with the current data it is
desirable that algorithms supported\footnote{While any number of
jet algorithms may in principle be included in an offline analysis
stream, in practice only a few algorithms will typically be fully
supported by detailed energy scale, resolution, and efficiency
corrections.} for the new data produce similar results.

Details in the D\O\ Run~I jet algorithm forced the splitting and
merging of jets to occur as they are found. In effect this defines
an order dependence based on the seed $E_T$ of the jets. It was
possible to test two orderings in the jet clustering. In the first
case, jets were initially found around all seed towers above a
1\,GeV threshold, then around all midpoints. In the second case
they were first found around all midpoints between seed towers,
then around the seed towers themselves. Fig.~\ref{CSRAT} shows
the $E_T$ distributions for three trials, the legacy seed, seed +
midpoint, and midpoint + seed trials. Also shown are the ratios of
the $E_T$ spectra. A cone radius of 0.7 was used.

\begin{figure}[htb]
\begin{center}
\epsfig{figure=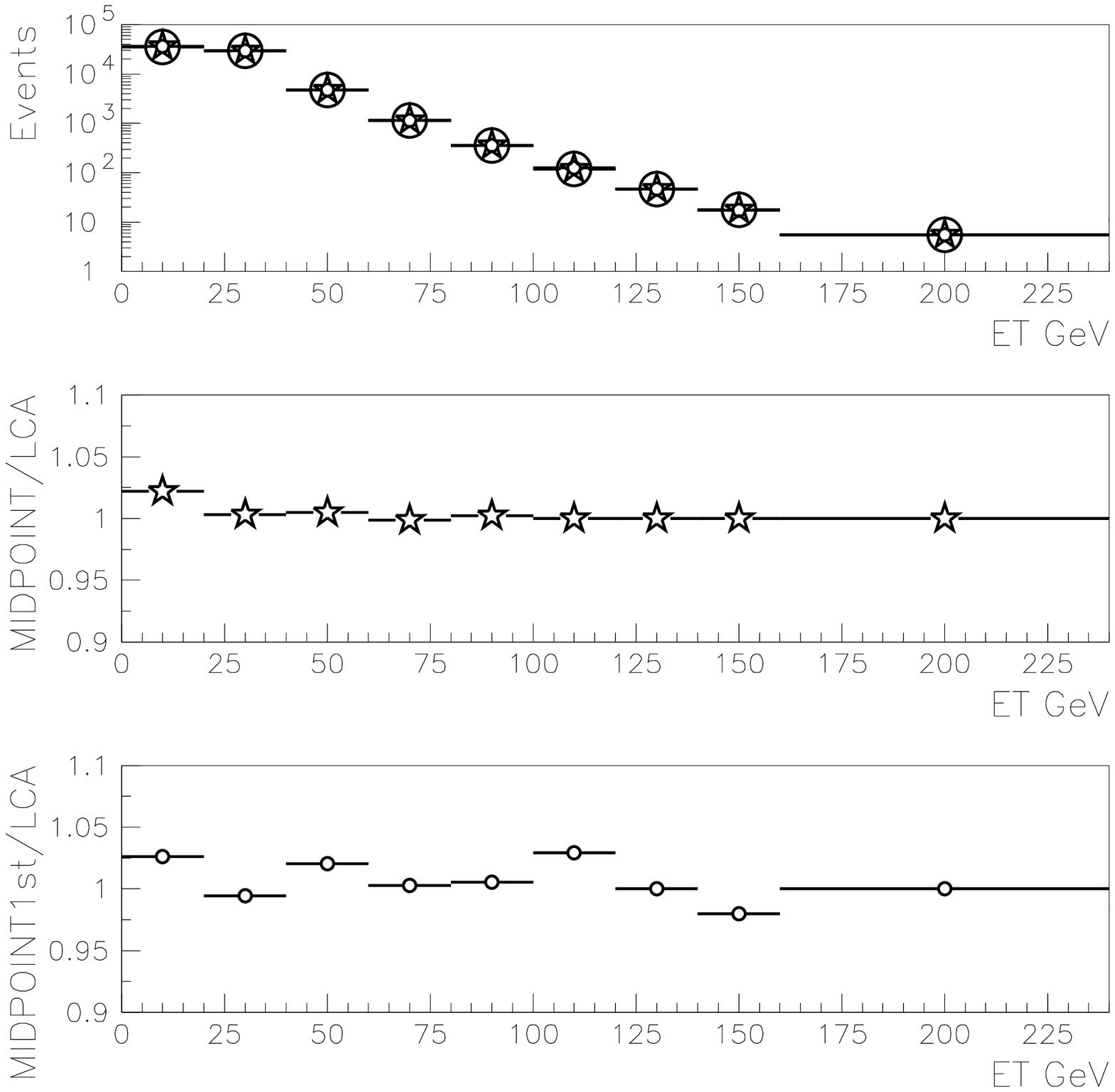,height=75mm}
\end{center}
\caption{Jet $E_T$ distributions and ratios. Top: Jet $E_T$ distributions
for the three algorithms overlayed. Legacy seeds (large circles), seeds +
midpoints (stars), midpoints + seeds (small circles).
Middle: Seeds + midpoint
distribution divided by the legacy distribution. Bottom: Midpoint + seeds
distribution divided by the legacy distribution.}
\label{CSRAT}
\end{figure}

There are two effects to observe in Fig.~\ref{CSRAT}. First, the addition of
midpoints tends to cause an increase in the number of low $E_T$ jets. This
is because the midpoints are effectively zero threshold seeds, therefore
very soft jets that tend to fail reconstruction by falling short of the seed
requirement may sometimes be reconstructed around a midpoint. Second, the
results are different depending on the order in which the seeds + midpoints
are used. However, we can safely conclude that the addition of midpoints has
little more than a few percent effect on the experimental jet $E_T$
distribution.

Fig.~\ref{ETRAT} shows the ratio of the leading jet for the legacy seed
and midpoint + seed algorithms. Since a meaningful test requires the
comparison of the same jets, the jets were also required to be matched
within a radius of 0.2 (in $\Delta\eta\times\Delta\phi$) to prevent
accidental comparisons of unrelated jets due to `flipping' of the jet order
between algorithms. Fig.~\ref{SMFRAC} shows the fractions of isolated,
merged, split, and multiply split/merged jets for the legacy seed and
midpoint + seed algorithms. In each case only small variations are observed
between the two algorithms, indicating that a legacy cone algorithm
augmented by midpoints is an acceptable choice for comparisons to Run~I
physics results. In fact, Figs.~\ref{CSRAT} and~\ref{ETRAT} represent
extreme deviations in jet $E_T$, since $E_T$ differences are expected to be
reduced after application of jet energy corrections appropriate to each
algorithm.

\begin{figure}[tbp]
\begin{minipage}[t]{75mm}
    \epsfig{figure=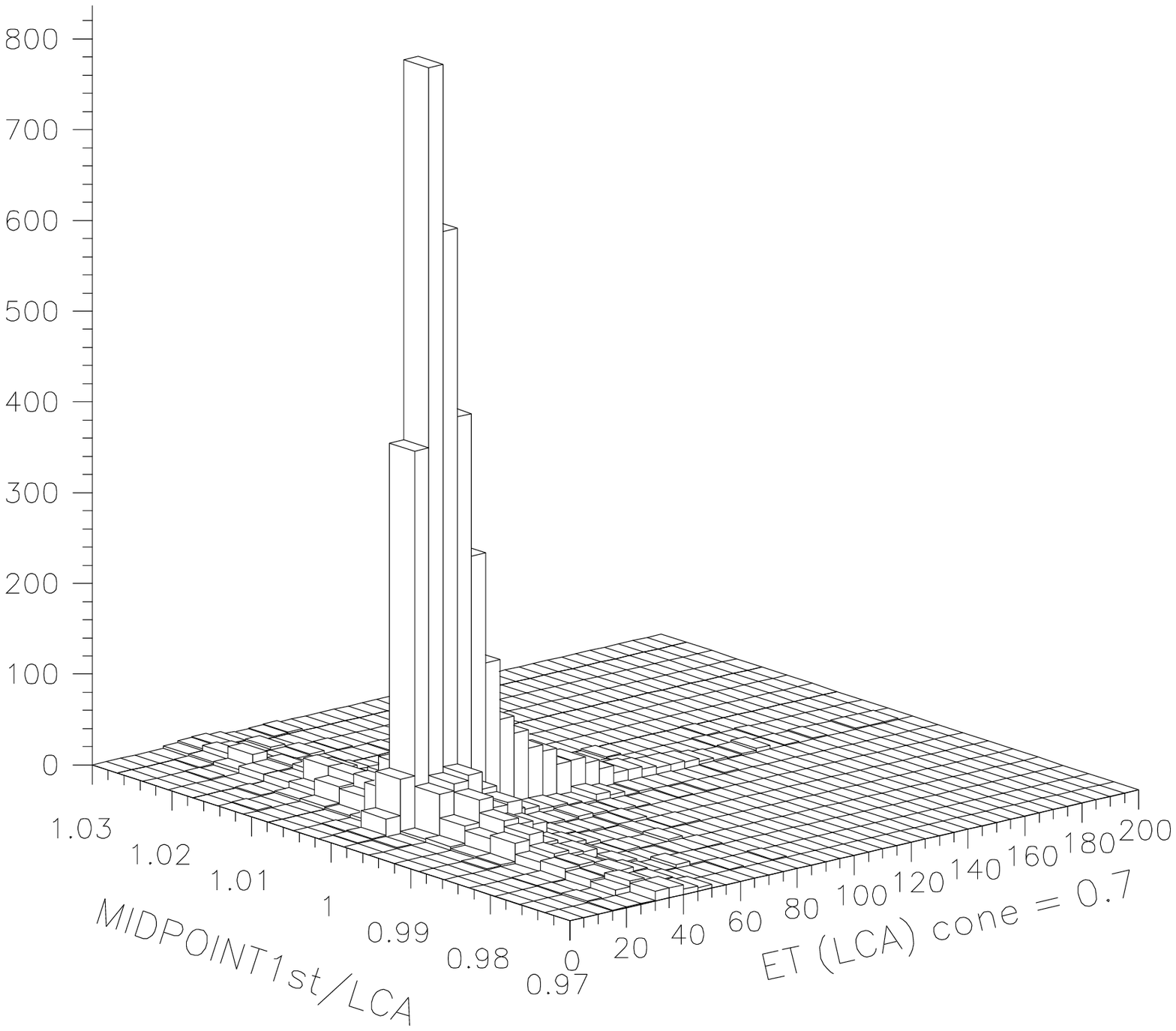,height=75mm}
    \caption{$E_T$ ratios for leading jets.  The ratio of leading jet $E_T$
    in the midpoint algorithm is plotted as a function of the legacy cone
    jet's $E_T$.}
    \label{ETRAT}
  \end{minipage}
\hspace{9mm}%
\begin{minipage}[t]{75mm}
    \epsfig{figure=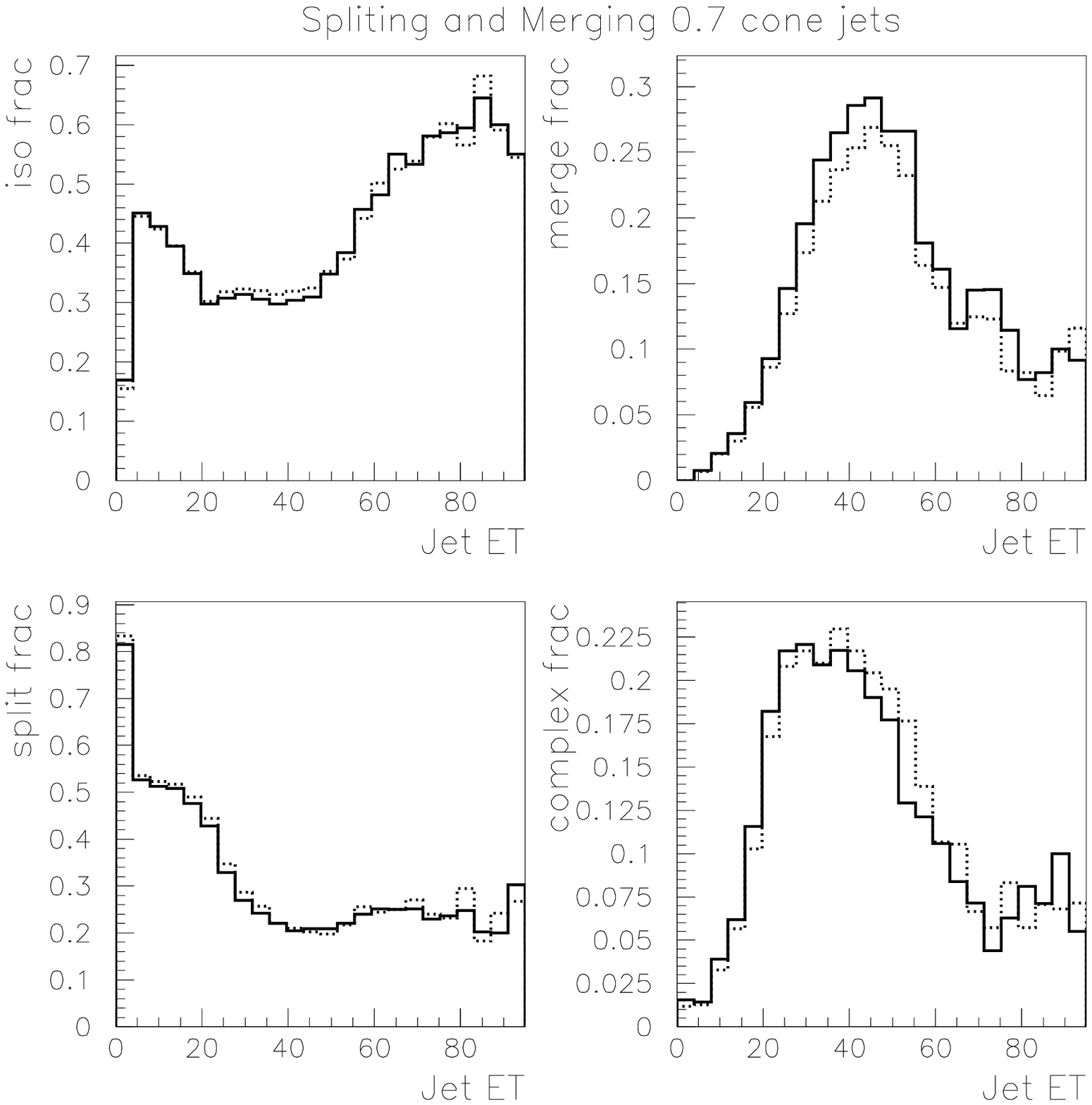,height=75mm}
    \caption{A view of splitting and merging fractions in the legacy seed
             (solid) and midpoint + seed algorithms (dotted).}
    \label{SMFRAC}
  \end{minipage}
\end{figure}

\subsection{Proposals for Common Run II Cone Jet Algorithms}

\label{ss_Prop}

The cone algorithm starts with a cone defined in E-scheme variables as
\begin{equation}
i\subset C\quad  : \quad\sqrt{\left( y^{i}-y^{C}\right)^{2}
+\left( \phi ^{i}-\phi ^{C}\right) ^{2}}\leq R.
\end{equation}
where for massless towers, particles, or partons $y^i = \eta^i$.
The E-scheme centroid corresponding to this cone is given by
\begin{eqnarray}
p^{C} &=&({E}^{C},{\bf {p}}^{C})=\sum_{i\subset
C}(E^{i},p_{x}^{i},p_{y}^{i},p_{z}^{i})\;, \\
\bar{y}^{C} &=&\frac{1}{2}\ln \frac{E^{C}+p_{z}^{C}}{E^{C}-p_{z}^{C}}\;,\quad
\bar{\phi}^{C}=\tan ^{-1}\frac{p_{y}^{C}}{p_{x}^{C}}\;.
\end{eqnarray}
A jet arises from a ``stable'' cone, for which $\bar{y}^{C}=y^C=y^J$ and
$\bar{\phi}^{C}=\phi^C=\phi^J$, and the jet has kinematic properties
\begin{eqnarray}
p^{J} &=&(E^{J},{\bf p}^{J})=\sum_{i\subset
J=C}(E^{i},p_{x}^{i},p_{y}^{i},p_{z}^{i})\;, \\
p_{T}^{J} &=&\sqrt{(p_{x}^{J})^{2}+(p_{y}^{J})^{2}}\;, \\
y^{J} &=&\frac{1}{2}\ln \frac{E^{J}+p_{z}^{J}}{E^{J}-p_{z}^{J}}\;,\quad
\phi ^{J}=\tan ^{-1}\frac{p_{y}^{J}}{p_{x}^{J}}\; .
\end{eqnarray}

{\em Seedless algorithm.} For a seedless algorithm we recommend the
streamlined jet algorithm defined in Section \ref{sec:seedless}
that includes the flow cut for computational efficiency improvement
and reduction of soft proto-jet construction.  The clustering or
jet finding should be done in terms of E-scheme variables.

{\em Seed--based algorithm or ILCA.} Backwards compatibility is
important here as well as common specifications between
experiments.  For the Run II algorithm we recommend that jet
clustering commence on each seed tower (rather than consolidated
seeds as in {\mbox Run~I}), for simplicity of the algorithm and to
reduce dependencies on detector segmentation.  Since the finding of
proto-jets is determined by the seed threshold, it is reasonable to
determine the midpoints based on the positions of the proto-jets
rather than the seed list itself, as illustrated in Fig.~\ref
{midpoint}. This would reduce the number of midpoints to be
calculated due to the large combinatorics caused by adjacent seed
towers within jet cones.

\begin{figure}[tbp]
\begin{center}
\epsfig{figure=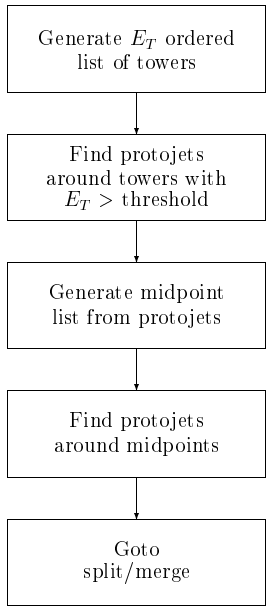,width = 35mm}
\end{center}
\caption{Method for addition of midpoints.}
\label{midpoint}
\end{figure}

{\em Specifications Summary} We list here the precise
specifications of the jet algorithms and variables:

\begin{enumerate}
\item  $R_{cone}$: 0.7

\item  $p_T^{seed}$: 1.0 GeV

\item  Recombination: E-scheme

\item  Midpoints: Added after cone clustering

\item  Split/Merge: $p_T$ ordered, threshold = 50\% of
lower $p_T$ jet

\item  Reported kinematic variables: E-scheme, either
directly as $(E^{J},{\bf p}^{J})$ or
as $(m^{J},{p_{T}}^{J},y^J,\phi^J)$, where $m^J$ is the mass
of the jet $(m^J=\sqrt{{E^J}^2-{{\bf p}^J}^2})$.
\end{enumerate}

\section{\KT Jet Algorithms}

\subsection{Introduction}

This section provides a guide for the definition of \KT jet
algorithms for the Tevatron. Section \ref{KT_DEF} describes the
recommended algorithm in detail. Section~\ref{subs:pre} discusses
preclustering of particles, cells, or towers for both the CDF and
\D0 experiments. Sections~\ref{subs:jes} and~\ref{KT_MOMRES}
outline momentum calibration of the \KT algorithm and briefly
describe jet resolution. Finally, in Section~\ref{KT_test}, we
provide a few examples of the versatility of the \KT algorithm.

\subsection{The Run II \KT Algorithm}
\label{KT_DEF}

In this section we propose a standard \KT jet algorithm for Run II
at the Fermilab Tevatron.  This proposal, based on studies of the
\KT algorithm by several groups~\cite{cat93,cat92,ell_sop},
establishes a common algorithm that satisfies the general criteria
presented in Section 1.

The \KT jet algorithm starts with a list of {\em preclusters} which
are formed from calorimeter cells, particles, or
partons.\footnote{Preclustering is discussed in detail in
Section~\ref{subs:pre}.}  Initially, each precluster is assigned a
vector
\begin{equation}
 (E,{\bf p}) = E \left( 1, \, \cos \phi \sin \theta, \, \sin \phi \sin \theta,
 \, \cos \theta \right)
\end{equation}
where $E$ is the energy associated with the precluster, $\phi$ is the azimuthal angle,
and $\theta$ is the polar angle with respect to the beam axis. For each precluster, we
calculate the square of the transverse momentum, $p_T^2$, using
\begin{equation}
 p_T^2 = p_x^2 + p_y^2
\end{equation}
and the rapidity, $y$, using\footnote{To avoid differences in the behavior
of the algorithm due to computational precision when $|y|$ is large, we
assign $y$ = $\pm$10 if $|y|$ $>$ 10.}
\begin{equation}
 y = \frac{1}{2} \ln \frac{E + p_z}{E - p_z}. \label{eq:y}
\end{equation}
A flowchart of the \KT algorithm is shown in
Fig.~\ref{fig:kt_flo}. Starting with a list of preclusters and an
empty list of jets, the steps of the algorithm are as follows:

\begin{enumerate}

\item For each precluster $i$ in the list, define
\begin{equation}
d_i = p_{T,i}^2 \;.
\end{equation}
For each pair $(i,j)$ of preclusters ($i$ $\neq$ $j$), define
\begin{eqnarray}
 \lefteqn{d_{ij}} && \!\!\! = \min\left(p_{T,i}^2, p_{T,j}^2 \right)
      \frac{ \Delta{\cal R}_{ij}^2} {D^2} \nonumber \\
 && \!\!\! = \min\left(p_{T,i}^2,p_{T,j}^2 \right)
      \frac{(y_i - y_j)^2 + (\phi_i - \phi_j)^2} {D^2} \label{eq:dij}
\end{eqnarray}
where $D \approx 1$ is a parameter of the jet algorithm.
For $D = 1$ and $\Delta{\cal R}_{ij} \ll 1$, $d_{ij}$ is the
minimal relative transverse momentum $k_{\perp}$ (squared) of one
vector with respect to the other.

\item Find the minimum of all the $d_i$ and $d_{ij}$ and label it $d_{min}$.

\item If $d_{min}$ is a $d_{ij}$, remove preclusters $i$ and $j$ from the
list and replace them with a new, merged precluster $(E_{ij},{\bf p}_{ij})$
given by
\begin{eqnarray}
\lefteqn{E_{ij}} && \!\! = E_i + E_j \;, \label{eq:esum}  \\
\lefteqn{{\bf p}_{ij}} && \!\! = {\bf p}_i + {\bf p}_j\;. \label{eq:psum}
\end{eqnarray}

\item If $d_{min}$ is a $d_i$, the corresponding precluster $i$ is
``not mergable.'' Remove it from the list of preclusters and add it to
the list of jets.

\item If any preclusters remain, go to step 1.

\end{enumerate}

The algorithm produces a list of jets, each separated by
$\Delta{\cal R} > D$. Fig.~\ref{fig:kt_example} illustrates how
the \KT algorithm successively merges the preclusters in a
simplified diagram of a hadron collision.

\begin{figure}[htb]
{\centerline{\psfig{figure=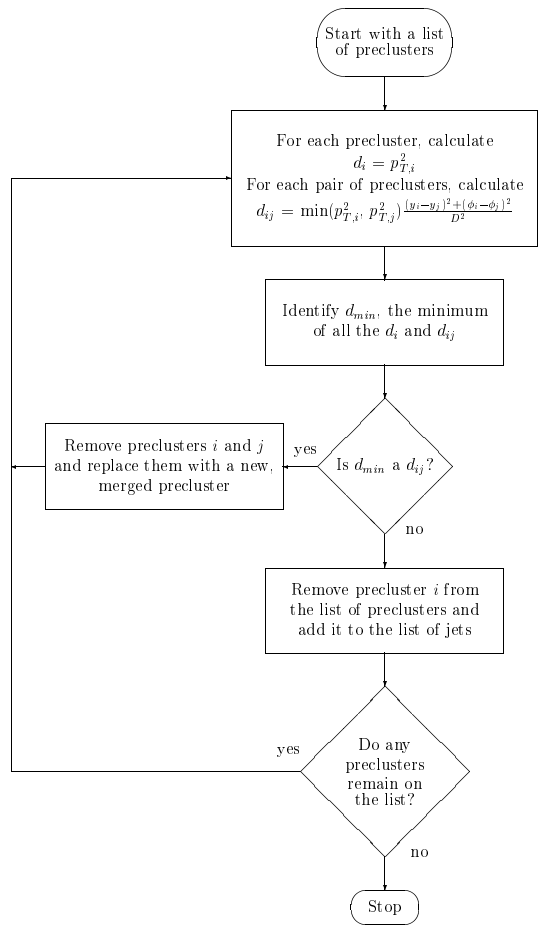,width=4.4in}}}
\caption{The \KT jet algorithm.}\label{fig:kt_flo}
\end{figure}

\begin{figure}[htb]
{\centerline{\psfig{figure=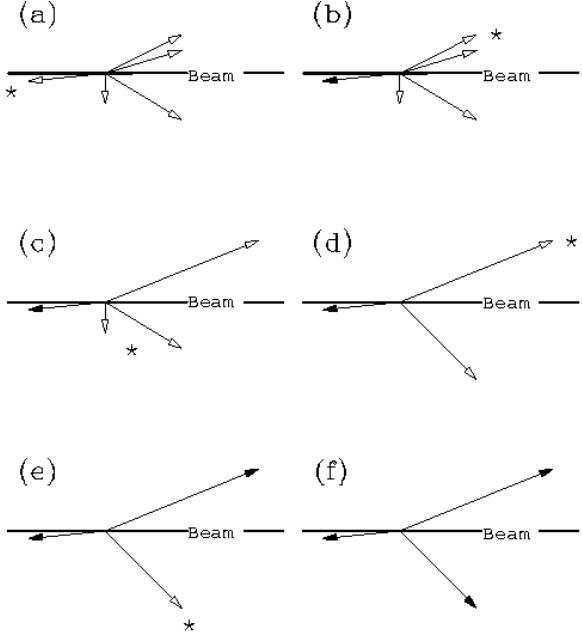,width=3.05in}}}
\caption{A simplified example of the final state of a hadron
collision.  The open arrows represent preclusters in the event, and
the solid arrows represent the final jets reconstructed by the \KT
algorithm. The six diagrams show successive iterations of the
algorithm. In each diagram, either a jet is defined (when it is
well separated from all other preclusters), or two preclusters are
merged (when they have small relative $k_{\perp}$). The asterisk
labels the relevant precluster(s) at each
step.}\label{fig:kt_example}
\end{figure}

The \KT algorithm presented above is based on several slightly
different \KT jet clustering algorithms for hadron
colliders~\cite{cat93,cat92,ell_sop}.  The main differences have to do with
(1) the recombination scheme and (2) the method of terminating the
clustering. The choices in the proposal above are discussed in the
following paragraphs.

The recombination scheme was investigated by
Catani {\em et al.}~\cite{cat93}.  We elect to use the covariant $E$-scheme
(Eqs.~\ref{eq:esum}--\ref{eq:psum}), which corresponds to vector
addition of four-momenta, because our goals are
\begin{enumerate}
\item conceptual simplicity,
\item correspondence to the scheme used in the \KT algorithm
      for $e^+e^-$ collisions~\cite{aleph},
\item absence of an energy defect~\cite{kosower}, and
\item optimum suitability for the calibration method
      described in Section~\ref{subs:jes}. \cite{frame}
\end{enumerate}

The prescription of Catani, {\em et al.}~\cite{cat93,cat92}
introduces a stopping parameter, $d_{cut}$, that defines the hard
scale of the physics process and separates the event into a hard
scattering part and a low-$p_T$ part (``beam jets'').  Catani {\em
et al.} suggest two ways to use the $d_{cut}$ parameter. First,
$d_{cut}$ can be set to a constant value {\it a priori}, and when
$d_{min}$ $>$ $d_{cut}$ the algorithm stops.  At this point, all
previously identified jets with $p^2_T$ $<$ $d_{cut}$ are
classified as beam jets, and all remaining preclusters with
$p^2_{T,i}$ $>$ $d_{cut}$ are retained as hard final-state jets.
Alternatively, an effective $d_{cut}$ can be identified on an
event-by-event basis so that clustering continues until a given number of
final-state jets are reconstructed.

Unlike Catani, {\em et al.}, the algorithm proposed by Ellis and
Soper~\cite{ell_sop} continues to merge preclusters until all jets are separated
by $\Delta{\cal R} > D$.  We have adopted this choice.  Besides its
simplicity, this method maintains a similarity with cone algorithms in
hadron collisions.  Whereas the use of $d_{cut}$ is well suited for defining
an {\em exclusive} jet cross section (typical of $e^+e^-$ collisions), we
desire an algorithm that defines {\em inclusive} jet cross sections in terms
of a single angular resolution parameter $D$, which is similar to $R$ for
cone algorithms.

\subsection{Preclustering}
\label{subs:pre}

As described in the previous section, the input to the \KT jet
algorithm is a list of vectors, or preclusters.  Ideally, one
should be able to apply the
\KT algorithm equally at the parton, particle, and detector levels, with no
dependence on detector cell type, number of cells, or size.  The goal of {\em
preclustering} is to strive for order independence and detector independence
by employing well-defined procedures to remove (or reduce) the
detector-dependent aspects of jet clustering.  Practically, however, this
independence is very difficult to achieve.  For example, if a single
particle strikes the boundary between two calorimeter towers, two clusters
of energy may be measured.  Conversely, two collinear particles may shower
in a single calorimeter tower so that only one vector is measured
experimentally.  Preclustering all vectors within a radius larger than
the calorimeter tower size removes this problem.

At the parton and particle levels, the simplest possible
preclustering scheme is to identify each parton or particle
four-vector as a precluster. Experimentally, differences between
the geometries of the CDF and D\O\ calorimeters necessitate
different preclustering schemes.  In particular, the D\O\
discussion describes how the preclustering scheme can be used to
control the number of preclusters passed to the \KT algorithm in
order to keep the jet analysis computationally feasible.  It can
also be used to ensure that the preclusters all exhibit positive
energy.  Candidate schemes to achieve these goals are described in
detail in the following sections.  However, it is important that
the preclustering scheme does not introduce the sort of problems
with infrared or collinear sensitivities that we earlier discussed
for the case of seeds.

\subsubsection{CDF Preclustering}

The CDF calorimeter system for Run II~\cite{cdf_runii} consists of 1,536
towers. Each tower consists of an electromagnetic (EM) component and a
hadronic (HAD) component.  In order to form preclusters for input to the
\KT algorithm, we propose the following:
\begin{enumerate}
\item Measure the amount of EM energy deposited
into each calorimeter tower, $E_{EM}$, and form the vector
($E_{EM},{\bf p}_{EM}$) where
\begin{eqnarray}
\lefteqn{p_{x,EM}} && \;\;\; = E_{EM} \cos \phi \sin \theta_{EM}\;, \\
\lefteqn{p_{y,EM}} && \;\;\; = E_{EM} \sin \phi \sin \theta_{EM}\;, \\
\lefteqn{p_{z,EM}} && \;\;\; = E_{EM} \cos \theta_{EM}\;.
\end{eqnarray}
Likewise, measure the amount of HAD energy deposited
into each calorimeter tower, $E_{HAD}$, and form the vector
($E_{HAD},{\bf p}_{HAD}$) where
\begin{eqnarray}
\lefteqn{p_{x,HAD}} && \;\;\;\;\; = E_{HAD} \cos \phi \sin \theta_{HAD}\;, \\
\lefteqn{p_{y,HAD}} && \;\;\;\;\; = E_{HAD} \sin \phi \sin \theta_{HAD}\;, \\
\lefteqn{p_{z,HAD}} && \;\;\;\;\; = E_{HAD} \cos \theta_{HAD}\;.
\end{eqnarray}
The angles $\theta_{EM}$, $\theta_{HAD}$ and $\phi$ specify the position
of the
calorimeter tower components with respect to the interaction point.
Note that $\theta_{EM}$ and $\theta_{HAD}$ may take on slightly different
values when calculated using different interaction points along the
beam axis (see Fig.~\ref{fig:cdf_caltow}).
\item For each calorimeter tower, calculate a vector ($E,{\bf p}$) by
summing the vectors for the EM and HAD components:
\begin{equation}
(E,{\bf p}) = (E_{EM}+E_{HAD}, \, {\bf p}_{EM} + {\bf p}_{HAD})
\end{equation}
\item For each calorimeter tower, calculate the $p_T$ from its
associated vector using
\begin{eqnarray}
\lefteqn{p_T} && \!\!\! = \sqrt{p_x^2 + p_y^2} \nonumber \\
 && \!\!\! = E_{EM} \sin \theta_{EM} + E_{HAD} \sin \theta_{HAD}\;.
\end{eqnarray}
\item Assemble a list of tower vectors for which
\begin{equation}
p_T > p_T^{min},
\end{equation}
where $p_T^{min} \approx$ 100~MeV.\footnote{This $p_T$ cut is
designed to retain towers with energies well above the level of
electronic noise. The exact value for this $p_T$ cut will depend on
measurements of calorimeter performance.}  These are the
preclusters for the \KT algorithm.
\end{enumerate}

In designing the CDF preclustering scheme, the primary goal was
simplicity.  We made every attempt to maintain a close relationship
between the physical calorimeter towers and the input preclusters for the
\KT algorithm.

\begin{figure}[htb]
{\centerline{\psfig{figure=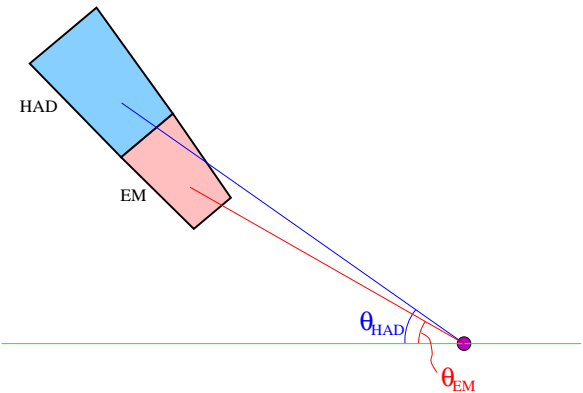,width=3.05in}}}
\caption{Schematic of a single CDF calorimeter
tower.}\label{fig:cdf_caltow}
\end{figure}

\subsubsection{D\O\ Preclustering}
The \KT jet algorithm is an ${\cal O}(n^3)$ algorithm, where $n$ is
the number of vectors in the event~\cite{cat93}. Limited computer
processing time does not allow this algorithm to run on the $\sim
\! 45000$ cells or even the $\sim \! 6000$ towers of the D\O\
calorimeter. Therefore, we employ a preclustering algorithm to
reduce the number of vectors input to the algorithm. Essentially,
towers are merged if they are close together in $\eta \times \phi$
space, or if they have small $p_T$ (or negative $p_T$, as explained
below). The preclustering algorithm described below was used by the
D\O\ experiment in Run I. We examine the effects of the Run I
preclustering algorithm, and discuss possible alternatives for Run
II. Although the effects of preclustering on jet observables should
be small, this is analysis and detector dependent. A Monte Carlo
study of preclustering effects on the jet $p_T$ and on jet
structure is also presented.

%
In Run I, one use of preclustering was to account for negative
energy calorimeter towers~\cite{jes_nim} which can cause
difficulties for the \KT algorithm. In the D\O\ calorimeter, we
measured the difference in voltage between two readings (peak minus
base), as illustrated in Fig.~\ref{fig:cal_signal2}. To first
order, this online baseline subtraction technique removes the
effect of luminosity-dependent noise in the calorimeter, on a
tower-by-tower basis. Residual fluctuations in each reading,
however, sometimes lead to measured energies which are negative.
One can imagine at least four ways to deal with these negative energy
towers.
\begin{enumerate}
\item Absorb the negative energy into a precluster of towers
such that the overall precluster energy is positive, as will be discussed
here.
\item Add an offset to all tower energies so that there are
none with negative energy.  The offset could then be removed later in the
analysis.
\item Ignore all towers with negative energy, {\it i.e.}, remove
them from the jet analysis.
\item Proceed with the \KT algorithm analysis including the negative
energy towers, assuming that their impact is negligible.  Recall that
in the cone algorithm case the negative energy towers are the source
of the observed limit cycles for quasi-stable cones, which does not seem to
be a serious problem.
\end{enumerate}
Clearly, further studies of this issue are required.
The precluster scheme can also be used to absorb low $p_T$ towers
similarly to what is done for negative energy towers.

\begin{figure}[htb]
\vskip-1cm
\begin{minipage}[t]{3.05in}
{\centerline{\psfig{figure=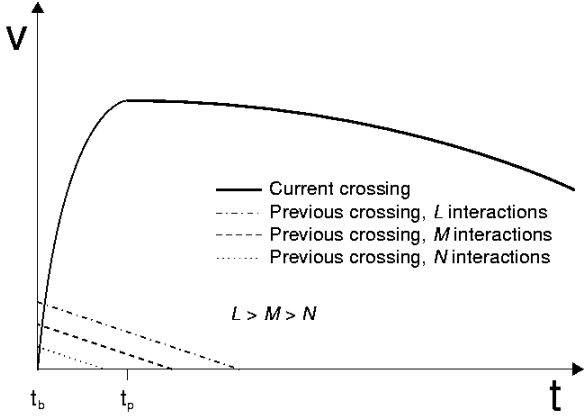,width=3.05in}}}
\vskip-1cm
\caption{
Schematic of voltage in a calorimeter cell as a function of time.
The solid line shows the contribution for a given event (the
current crossing). The cell is sampled once at $t_b$, just before a
$p\bar{p}$ bunch crossing, to establish a base voltage. The voltage
rises during the time it takes electrons to drift in the liquid
argon gap ($\sim$500 ns), and reaches a peak value at $t_p \approx
2 \, \mu\rm{s}$, which is set by pulse-shaping amplifiers in the
signal path. The cell is sampled again at $t_p$, and the voltage
difference $\Delta V = V(t_p) - V(t_b)$ is proportional to the raw
energy in the cell. Because the decay time of the signal $\tau
\approx 30 \, \mu\rm{s}$ is much larger than the accelerator bunch
crossing time $t_x = 3.5 \, \mu\rm{s}$, $V(t_b)$ may have a
contribution from a previous bunch crossing. The size of this
contribution is related to the number of $p\bar{p}$ interactions in
the previous crossing, which depends on the beam luminosity. The
dashed lines show an example contribution from a previous bunch
crossing containing three different numbers of $p\bar{p}$
interactions. The figure is not drawn to scale.}
  \label{fig:cal_signal2}
\end{minipage}
\hspace*{2mm}
\vskip-0.5cm
\end{figure}

The Run I preclustering algorithm, which is employed in the following
studies, has six steps:

\begin{enumerate}
\item Identify each calorimeter cell with a 4-vector
$(E,{\bf p}) =E \left( 1, \cos \phi \sin \theta, \sin \phi \sin
\theta, \cos \theta \right)$ where $E$ is the measured energy
in the cell.  For each cell, define
\begin{equation}
p_{T} = \sqrt{p_{x}^{2} + p_{y}^{2}} = E \sin \theta \\
\end{equation}
and
\begin{equation}
\eta = - \ln \left( \tan \frac{\theta}{2} \right)\; .
\end{equation}

\item Remove any calorimeter cells with $p_T < -500$ MeV.
Cells with slightly negative $p_T$ are allowed due to pileup
effects in the calorimeter, but cells with highly negative $p_T$
are very rarely observed in minimum bias events and are thus
considered spurious, so they are removed.

\item For each calorimeter tower, sum the transverse momenta
of cells within that tower:
\begin{equation}
p_{T}^{tower} = \sum_{cell \, \in \, tower} p_{T}^{cell}\; .
\end{equation}

\item
Merge towers if they are close together in $\eta \times \phi$
space:

\begin{enumerate}

\item
Form an $\eta$-ordered (from most negative to most positive) list
of towers; towers with equal $\eta$ are ordered from $\phi = 0$ to
$\phi = 2 \pi$.

\item
\label{it:top}
Remove the first tower in the list and call it a precluster.

\item
\label{it:closest}
From the remainder of the list, find the closest tower
to the precluster.

\item
\label{it:combine}
If they are within $\Delta{\cal R}_p = \sqrt{ \Delta \eta^2 +
\Delta \phi^2} = 0.2$, remove the closest tower from the list,
and combine it with the existing precluster, forming a new
precluster; go to~\ref{it:closest}.

\item
If any towers remain, go to~\ref{it:top}.

\end{enumerate}

\item Preclusters which have negative transverse momentum
$p_T = p_{T-} < 0$ are redistributed to neighboring preclusters.
Given a negative $p_T$ precluster with
$(p_{T-},\eta_{-},\phi_{-})$, we define a search square of size
$(\eta_{-} \pm 0.1) \times (\phi_{-} \pm 0.1)$. If the vector sum
of positive $p_T$ in the search square is greater than $|p_{T-}|$,
then $p_{T-}$ is redistributed to the positive $p_T$ preclusters in
the search square. Otherwise, the search square is increased in
steps of $\Delta \eta = \pm 0.1$ and $\Delta \phi = \pm 0.1$, and
redistribution is again attempted. If redistribution still fails
with a search square of size $(\eta_{-} \pm 0.7) \times (\phi_{-}
\pm 0.7)$, the $p_T$ of the negative momentum precluster is set to zero.

\item Preclusters which have $p_T < p_T^p = 200$ MeV are
redistributed to neighboring preclusters, as in step 5.  We make
the additional requirement that the search square have at least
three positive $p_T$ preclusters, to reduce the overall number of
preclusters. The threshold $p_T^p$ was tuned to produce $\sim \!
200$ preclusters/event, as shown in Fig.~\ref{fig:ktcl_page6}, to
fit processing time constraints. Next, jets are reconstructed from
the preclusters.

\end{enumerate}
In steps 4--6, the combination followed a Snowmass style
prescription:
\begin{eqnarray}
p_T &=& p_{T,i} + p_{T,j} \;, \label{pt1}\\
\eta &=& \frac{p_{T,i}\eta_i + p_{T,j}\eta_j} {p_{T,i} + p_{T,j}} \;, \\
\phi &=& \frac{p_{T,i}\phi_i + p_{T,j}\phi_j} {p_{T,i} + p_{T,j}} \;.
\end{eqnarray}

As a minimal change to the Run I preclustering algorithm, a
possible Run II preclustering proposal should instead use vector
addition of four-momenta. The Run II preclustering
algorithm should also use $y$ (as defined by Eq.~\ref{eq:y}) instead of
$\eta$ and a true 2-vector $p_T$ rather than the scalar $p_T$ of
Eq.~\ref{pt1}.  Generally, the definitions of variables and recombination
scheme in the preclustering algorithm should match the choices used
in the proposed \KT jet algorithm. All of the results presented
here used the Run I preclustering algorithm.

\vspace{5mm}
\begin{figure}[htb]
\vskip-1cm
\begin{minipage}[t]{3.05in}
{\centerline{\psfig{figure=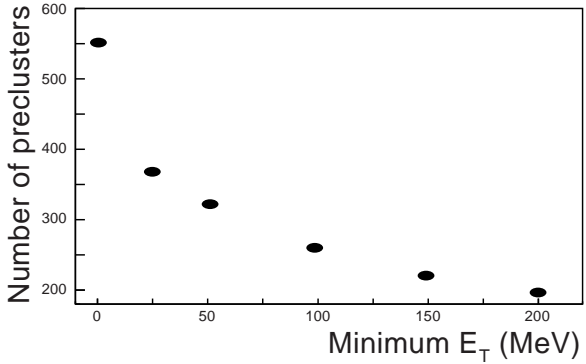,width=3.05in}}}
\vskip-1cm
\caption{
The number of preclusters per event, as a function of minimum
precluster transverse energy $E_T^p$. The D\O\ data were
preclustered with the choice $E_T^p = 200$ MeV, which produced
$\sim$200 preclusters per event. With the preclusters treated as massless,
$E_T$ is the same as $p_T$.  This identification is certainly appropriate for
individual calorimeter towers.}
  \label{fig:ktcl_page6}
\end{minipage}
\hspace*{2mm}
\vskip-0.5cm
\end{figure}

The preclustering radius $\Delta{\cal R}_p$ in step 4 of the
algorithm above can be used to test the sensitivity of jets to the
calorimeter segmentation size, $\Delta \phi \times \Delta \eta
= 0.1 \times 0.1$ (or smaller) in the D\O\ calorimeter.
Preclustering with $\Delta{\cal R}_p = 0.2
> \Delta \eta$ or $\Delta \phi$ in step 4 of the algorithm
mimics a coarser calorimeter. This effect was studied in a sample
of {\sc herwig} Monte Carlo QCD jet events. The jets in the hard
$2\rightarrow2$ scattering were generated with $p_T > 50$ GeV, and
at least one of the two leading order partons was required to be
central ($|\eta| < 0.9$). The events were passed through a full
simulation (including luminosity ${\cal L} \approx 5 \times 10^{30}
\rm{cm}^{-2}\rm{s}^{-1}$) of the D\O\ detector. The MC sample is
described in more detail in Section~\ref{sec:off}.
Fig.~\ref{fig:nopre_cl_1} shows the number of preclusters with
$\Delta{\cal R}_p = 0.2$ is $\sim$180, reduced by $37 \%$ from that
obtained with $\Delta{\cal R}_p = 0$.  Fig.~\ref{fig:nopre_cl_2}
shows that preclustering is necessary even at the particle level in
the Monte Carlo, reducing the number preclusters by $24 \%$.
Comparing Figs.~\ref{fig:nopre_cl_1} and \ref{fig:nopre_cl_2}, the
number of preclusters in the detailed detector simulation is a
factor 2.4 higher than at the particle level for $\Delta{\cal R}_p
= 0$. Most of the additional preclusters are reconstructed near the
beampipe and some are due to localized deposits of low energy. With
$\Delta{\cal R}_p = 0.2$, the number of preclusters increases only
by a factor 2.0.

\begin{figure}[htb]
\centerline{\psfig{figure=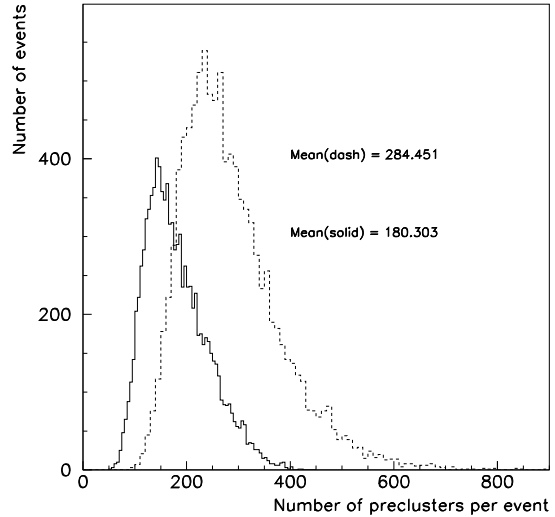,width=3.05in}}
\caption{Distribution of the number of preclusters per event,
with $\Delta{\cal R}_p = 0.2$ (solid), and with $\Delta{\cal R}_p =
0$ (dash). Taken from a sample of QCD jet events from MC data. The
jets were reconstructed using the calorimeter simulation, including
the luminosity simulation. The preclustering radius $\Delta{\cal
R}_p = 0.2$ reduces the mean number of preclusters per event by $37
\%$. }
\label{fig:nopre_cl_1}
\end{figure}

\begin{figure}[htb]
\centerline{\psfig{figure=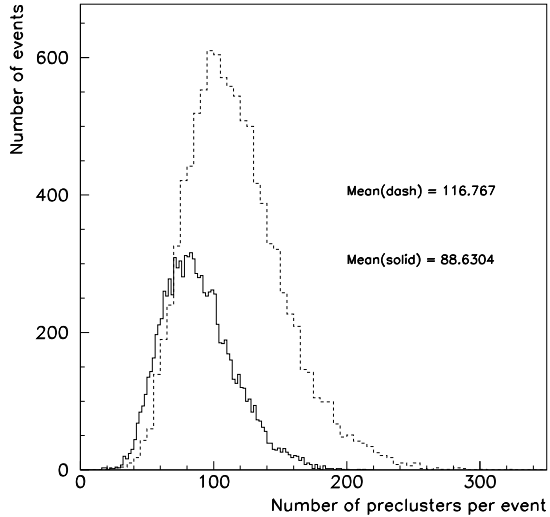,width=3.05in}}
\caption{
Same as in Fig.~\ref{fig:nopre_cl_1}, except the jets were
reconstructed in MC data at the particle level, with no calorimeter
or luminosity simulation. The same preclustering radius
$\Delta{\cal R}_p = 0.2$ reduces the mean number of preclusters per
event by $24 \%$}.
\label{fig:nopre_cl_2}
\end{figure}

The effect of the preclustering radius $\Delta{\cal R}_p$ on jets
and jet structure was examined next. Fig.~\ref{fig:nopre_et_1}
shows the comparison of the leading jet $p_T$ with $\Delta{\cal
R}_p = 0.2$ to that with $\Delta{\cal R}_p = 0$. The jets were
reconstructed with the \KT jet algorithm $D = 0.5$. The
preclustering radius $\Delta{\cal R}_p = 0.2$ (step 4 of the
preclustering algorithm) reduces the mean jet $p_T$ by 0.7 GeV.
Evidently, the preclustering algorithm assigns energy differently
than the \KT algorithm. It is difficult to track exactly which
towers end up in each jet, in part because of the redistribution of
energy in steps 5 and 6 of the preclustering algorithm. The net
effect is that some energy belonging to the leading jet when
$\Delta{\cal R}_p = 0$ is transferred to non-leading jets when
$\Delta{\cal R}_p = 0.2$. The shift in the leading jet $p_T$
spectrum is visible in the top panel of Fig.~\ref{fig:nopre_et_1},
and the ratio in the bottom panel suggests some dependence on the
jet $p_T$. Such a shift may need to be corrected for in the Run II
experimental data, but will be different due to the change in
calorimeter electronics. In Run I, a correction was not explicitly
applied to the experimental data for this effect. Instead, the
theoretical predictions included the identical preclustering
algorithm used in experimental data. Fortunately, the particle-level
result for leading jet $p_T$ is not as sensitive to
$\Delta{\cal R}_p$. This is shown in Fig.~\ref{fig:nopre_et_2}.
Note that even the particles in the Monte Carlo were projected into
a calorimeter-like grid ($\Delta \phi \times \Delta \eta$
= 0.1 $\times$ 0.1) by the preclustering algorithm. If
this were not the case, then we would expect an even larger effect
than illustrated in Fig.~\ref{fig:nopre_et_2}.

\begin{figure}[htb]
\centerline{\psfig{figure=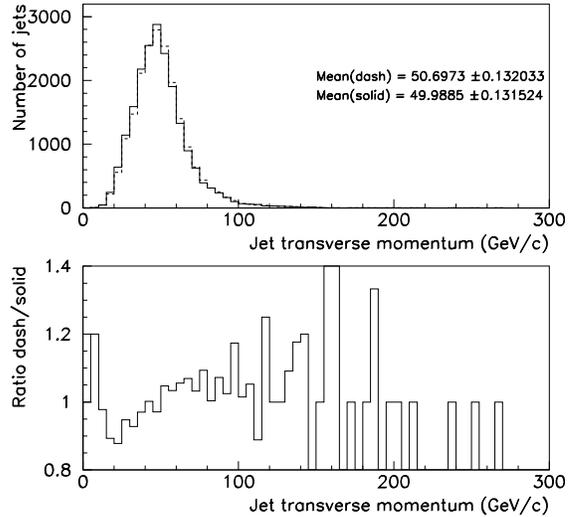,width=3.05in}}
\caption{The top panel shows the distribution of the leading jet $p_T$
with $\Delta{\cal R}_p = 0.2$ (solid), and with $\Delta{\cal R}_p =
0$ (dash). Measured in a sample of QCD jet events from MC data. The
sample was generated with minimum parton transverse momentum
$p_T^{min}$ = 50~GeV. The \KT jets were reconstructed with $D
= 0.5$ in the calorimeter simulation, including the luminosity
simulation. The preclustering radius $\Delta{\cal R}_p = 0.2$
reduces the mean of the leading jet $p_T$ by 0.7 GeV. The bottom
panel shows the ratio of the histograms in the top panel. }
\label{fig:nopre_et_1}
\end{figure}

\begin{figure}[htb]
\centerline{\psfig{figure=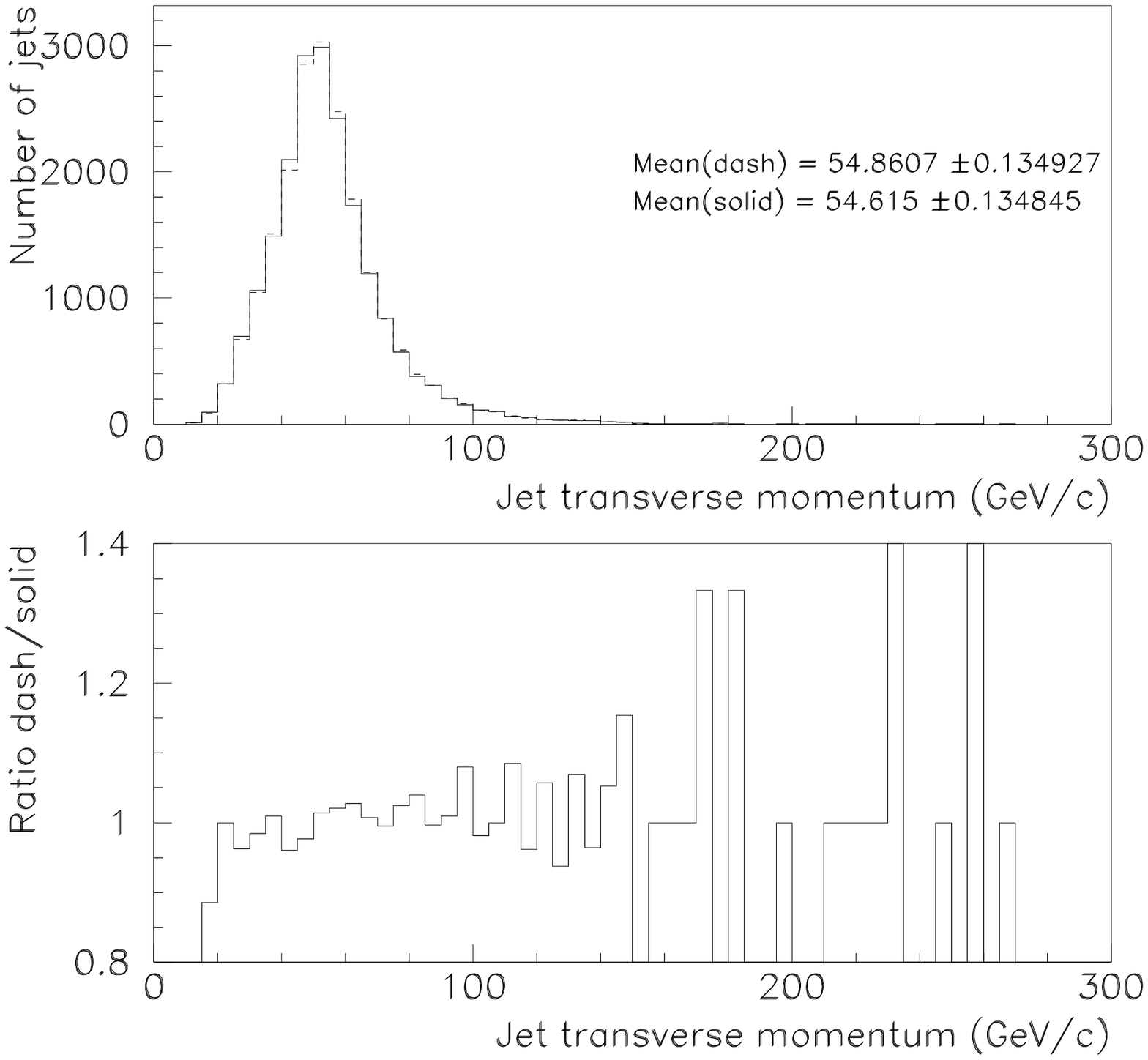,width=3.05in}}
\caption{
Same as in Fig.~\ref{fig:nopre_et_1}, except the jets were
reconstructed in MC data at the particle level, with no calorimeter
or luminosity simulation. The same preclustering radius
$\Delta{\cal R}_p = 0.2$ reduces the mean of the leading jet $p_T$
by 0.25 GeV. The bottom panel shows the ratio of the histograms in
the top panel. }
\label{fig:nopre_et_2}
\end{figure}

The jet structure, however, is more sensitive to the preclustering
radius $\Delta{\cal R}_p$. Fig.~\ref{fig:nopre_y_page2_bottom}
shows the average subjet multiplicity, as a function of $y_{cut}$
(see Section \ref{subs:structure}), in particle-level jets. There
are more subjets in jets when $\Delta{\cal R}_p = 0$, compared to
when $\Delta{\cal R}_p = 0.2$. Requiring preclusters to be
separated by $\Delta{\cal R}_p$ affects the subjet structure below
\begin{eqnarray}
y_{cut} & < & \left( \frac{\Delta{\cal R}_p} {2D} \right)^2 \nonumber \\
& < & 10^{-1.4}.
\end{eqnarray}
Again, the subjet multiplicity is increased even further
when particles in the Monte Carlo are not
projected into a calorimeter-like grid ($\Delta
\phi \times \Delta \eta$ = 0.1 $\times$ 0.1).
This underscores the fact that the same preclustering algorithm, as
well as the same jet algorithm, must be used in any comparisons of
theoretical predictions to experimental data which are sensitive to
internal jet structure at the level of the detector granularity.

\begin{figure}[htb]
\centerline{\psfig{figure=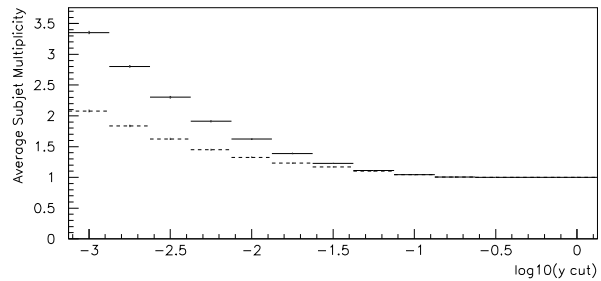,width=3.05in}}
\caption{
The average subjet multiplicity,
as a function of $y_{cut}$,
in a sample of jets reconstructed in MC data
at the particle level, with no calorimeter or luminosity simulation.
The solid curve shows the results with $\Delta{\cal R}_p = 0$,
and the dashed curve shows the results with $\Delta{\cal R}_p = 0.2$.
The preclustering radius $\Delta{\cal R}_p = 0.2$
reduces the average subjet multiplicity for $y_{cut} < 10^{-1.4}$.
}
\label{fig:nopre_y_page2_bottom}
\end{figure}

\subsection{Momentum Calibration of $K_{T}$ Jets at D\O\ }
\label{subs:jes}


Jet production is the dominant process in \ppbar\ collisions at
$\sqrt{s}$ = 1.8 TeV, and almost every physics measurement at the
Tevatron involves events with jets. A precise calibration of
measured jet momentum and energy, therefore, is of fundamental
importance. Although the use of a $K_{T}$ algorithm is well defined
theoretically, questions  have recently arisen regarding the
performance of the algorithm in a high
luminosity hadron collider environment.

The \D0 Collaboration developed a method to calibrate $K_{T}$ jets
to a high level of accuracy. The details are discussed thoroughly
in Ref.~\cite{calor99,ktnim}. Here, we briefly summarize this work
by the \D0 Collaboration to illustrate instrumentation effects on
the $K_{T}$ algorithm, as well as its behavior in a high luminosity
hadron collider. The $K_{T}$ jets momentum scale correction is
largely based on the calibration of cone jets, extensively
discussed in a recent article~\cite{jes_nim}. The derivation of the
momentum scale correction is performed for $K_{T}$ jets with $D=1$.
The measured jet momentum, $p_{jet}^{meas}$, is corrected to that
of the final-state particle-level jet, $p_{jet}^{ptcl}$, using the
following relation:
\begin{equation}
  p_{jet}^{ptcl} = \frac{p_{jet}^{meas}- p_{O}}
    {R_{jet}} \, ,
\end{equation}
where $p_O $ denotes a momentum offset correction for underlying
event, uranium noise, pile-up, and additional \ppbar\ interactions. $R_{jet}$
is the calorimeter momentum response to jets.
Note that the equation is missing the
out-of-cone showering loss factor. In cone jets, this factor corrects for
the fraction of the energy of the final-state hadrons which is lost
outside the cone boundaries due to calorimeter showering.
This is an instrumentation
effect completely unrelated to parton showering losses outside the cone.
There is no correction for the latter.  Note that the important issue
here is not so much that $p_{O}$ be small or that $R_{jet}$ be near unity,
but rather that these parameters can be determined with precision.  This
is the question to be addressed when comparing jet algorithms.

The D\O\ uranium-liquid argon sampling calorimeters~\cite{nim} are
shown in Fig.~\ref{fig:calorim}--\ref{fig:cal_side_eta}. They
constitute the primary system used to identify $e$, $\gamma$, jets
and missing transverse energy (\vmet). \vmet\ is defined as the
negative of the vector sum of the calorimeter cell transverse
energies ($E_{T}$'s). The Central (CC) and End (EC) Calorimeters
contain approximately seven and nine interaction lengths of
material respectively, ensuring containment of nearly all particles
except high \pt\ muons and neutrinos. The intercryostat region
(IC), between the CC and the EC calorimeters, is covered by a
scintillator based intercryostat detector (ICD) and massless gaps
(MG)~\cite{nim}.  The segmentation is $\Delta
\phi \times \Delta \eta$ = 0.1 $\times$ 0.1 (or smaller).

\begin{figure}[htb]
\vskip-1cm
\begin{minipage}[t]{3.05in}
{\centerline{\psfig{figure=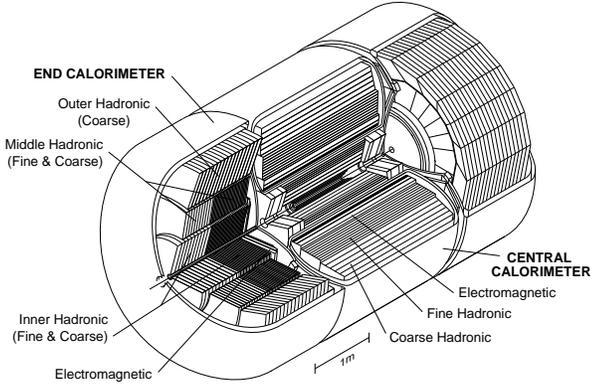,width=3.05in}}}
\vskip-1cm
\caption{
The D\O\ liquid argon calorimeter is divided physically into three
cryostats, defining the central calorimeter and two end
calorimeters. Plates of absorber material are immersed in the
liquid argon contained by the cryostats. Each cryostat is divided
into an electromagnetic, fine hadronic, and coarse hadronic
section. }
  \label{fig:calorim}
\end{minipage}
\hspace*{2mm}
\vskip-0.5cm
\end{figure}

\begin{figure}[htb]
\begin{minipage}[t]{3.05in}
{\centerline{\psfig{figure=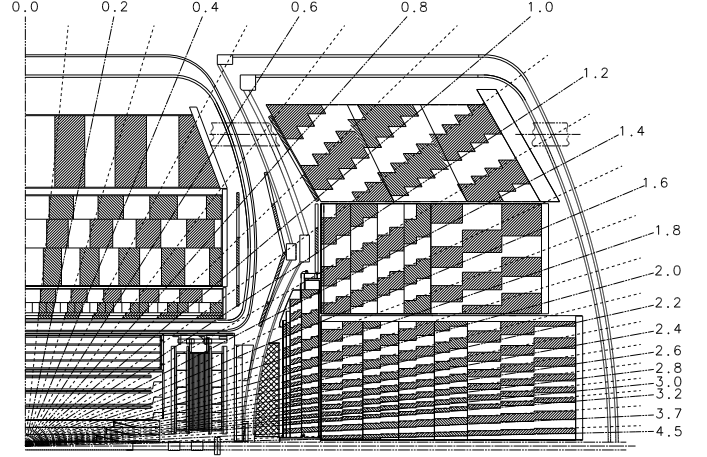,width=3.00in}}}
\vskip-1cm
\caption{
One quadrant of the D\O\ calorimeter and drift chamber,
projected in the $x - z$ plane.
Radial lines illustrate the detector pseudorapidity
and the pseudoprojective geometry of the calorimeter towers.
Each tower is of size
$\Delta \eta \times \Delta \phi = 0.1 \times 0.1$.
}
  \label{fig:cal_side_eta}
\end{minipage}
\hspace*{2mm}
\vskip-0.5cm
\end{figure}

The fractional energy resolution, $\sigma_{E} / E$, characterizes
the suitability of the D\O\ calorimeter system for {\it in-situ}
momentum calibration techniques.  It is parameterized with a
$\sqrt{S^2 /E + C^2}$ functional form.  For electrons, the sampling
term, $S$, is 14.8 (15.7)\% in the CC (EC), and the constant term,
C, is 0.3\% in both the CC and EC. For pions, the sampling term is
47.0 (44.6)\%, and the constant term is 4.5 (3.9)\% in the CC (EC).
The energy response is linear to within 0.5\% for electrons above
10 \gev\ and for pions above 20~GeV. The D\O\ calorimeters are
nearly compensating, with an $\frac{e}{\pi}$ ratio less than 1.05
above 30 GeV. Due to the hermiticity and linearity of the D\O\
calorimeters their response function is well described by a
Gaussian distribution. These properties indicate that the D\O\
calorimeter system is well suited for jet and \met\ measurements
and are the basis of the {\it in-situ} calibration method described
here.

\subsubsection{Offset Correction}\label{sec:off}

The total offset correction is measured in transverse momentum and
expressed as $p_{T,O} = O_{ue} + O_{zb}$. The first term is the
contribution of the underlying event (energy associated with the
spectator partons in a high $p_{T}$ event). The second term accounts
for uranium noise, pile-up and energy from additional \ppbar\
interactions in the same crossing.
Pile-up is the residual energy from previous \ppbar\
crossings as a result of the long shaping times associated with the
preamplification stage in calorimeter readout cells.

To simulate the contribution of $O_{zb}$ to jets, \D0 Run~I
collider data taken in a random \ppbar\ crossing (no trigger
requirements) was overlayed on high $p_T$ jet {\sc
herwig}~\cite{herwig} Monte Carlo events. Jets were matched in this
sample to jets in the sample with no overlay. The contribution of
uranium noise, pile-up, and multiple interactions was determined by
taking the difference in $p_T$ between matched pairs. $O_{ue}$ was
extracted in the same way from the overlap of low luminosity
minimum bias data (a crossing with an inelastic collision) on Monte Carlo
events. $O_{ue}$ and $O_{zb}$ for jets with $p_{T}=30-50$~GeV are
shown in Figs.~\ref{fig:oue} and \ref{fig:ozb}. The offset is derived
in the central calorimeter and extrapolated to higher $\eta$
regions.

\begin{figure}[htb]
\vskip0.5cm
\begin{minipage}[t]{3.05in}
{\centerline{\psfig{figure=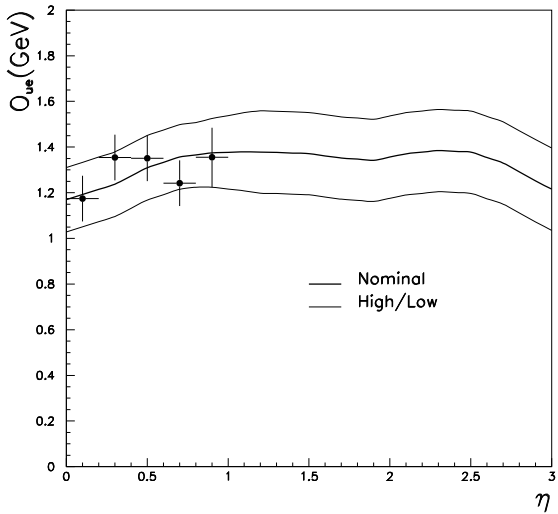,width=3.05in}}}
\vskip-3cm
\caption{Physics underlying event offset $O_{ue}$ versus $\eta$.
Above $\eta$ = 0.9, the result is an extrapolation.}
  \label{fig:oue}
\end{minipage}
\hspace*{2mm}
\vskip-0.5cm
\end{figure}

\begin{figure}[htb]
\vskip0.5cm
\begin{minipage}[t]{3.05in}
{\centerline{\psfig{figure=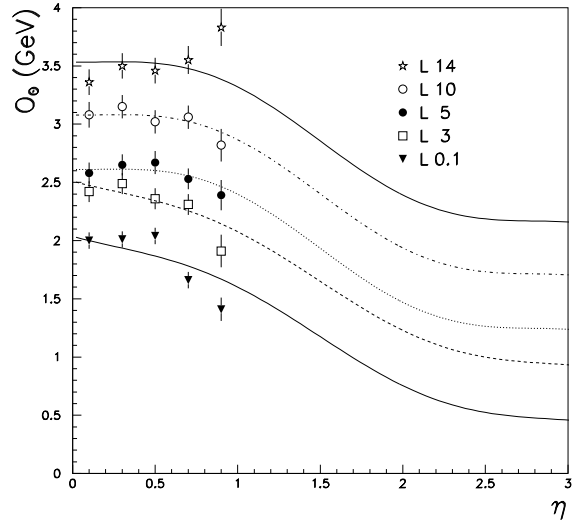,width=3.05in}}}
\vskip-1cm
\caption{Offset due to uranium noise, pile-up and multiple
         interactions, $O_{zb}$ versus $\eta$ for different luminosities
         in units of 10$^{30}$ cm$^{-2}$sec$^{-1}$. Above $\eta$ = 0.9, the
         result is an extrapolation.}
\label{fig:ozb}
\end{minipage}
\hspace*{2mm}
\vskip-0.5cm
\end{figure}

\subsubsection{Response: The Missing $E_{T}$ Projection
Fraction Method}\label{sec:mpf}

\D0 makes a direct measurement of the jet momentum response using
conservation of \pt\ in Run~I photon-jet ($\gamma$-jet)
collider events~\cite{jes_nim}. Previously, the photon energy/momentum
scale was
determined from the
\D0 $Z \rightarrow e^{+}e^{-}$, $J/\psi$ and
$\pi^{\circ}$ data samples, using the masses of these known
resonances.
In the case of a $\gamma$-jet two body process, the jet momentum response
can be measured as:
\begin{equation}
 R_{jet} =
  1 + \frac{\vmet \cdot \hat{n}_{T\gamma}}{p_{T\gamma}} \, ,
\end{equation}
where $p_{T\gamma}$ and $\hat{n}$ are the transverse
momentum and direction of the photon. To avoid resolution
and trigger biases, $R_{jet}$ is binned in terms of $E^{\prime} =
p_{T\gamma}^{meas}
\cdot {\rm cosh} (\eta_{jet})$ and then mapped onto
$p_{jet}^{meas}$. $E^{\prime}$ depends only on photon variables and
jet pseudorapidity, which are quantities measured with very good
precision. $R_{jet}$ and $E^{\prime}$ depend only on the jet
position, which has little dependence on the type of jet algorithm
employed.

$R_{jet}$ as a function of $p^{meas}_{jet}$ ($p_{Kt}$) is shown in
Fig.~\ref{fig:mom_dep}.  The data is fit with the functional form
$R_{jet}(P) = a + b \cdot {\rm ln}(P) + c \cdot {\rm ln}(P)^2$.
$R_{jet}$ for cone ($R=0.7$)~\cite{jes_nim} and $K_{T}$ ($D=1$)
jets are different by about 0.05. This difference does not have any
physical meaning; it arises from different voltage-to-energy
conversion factors at the cell level before reconstruction.

\begin{figure}[htb]
\vskip-1cm
\begin{minipage}[t]{3.05in}
{\centerline{\psfig{figure=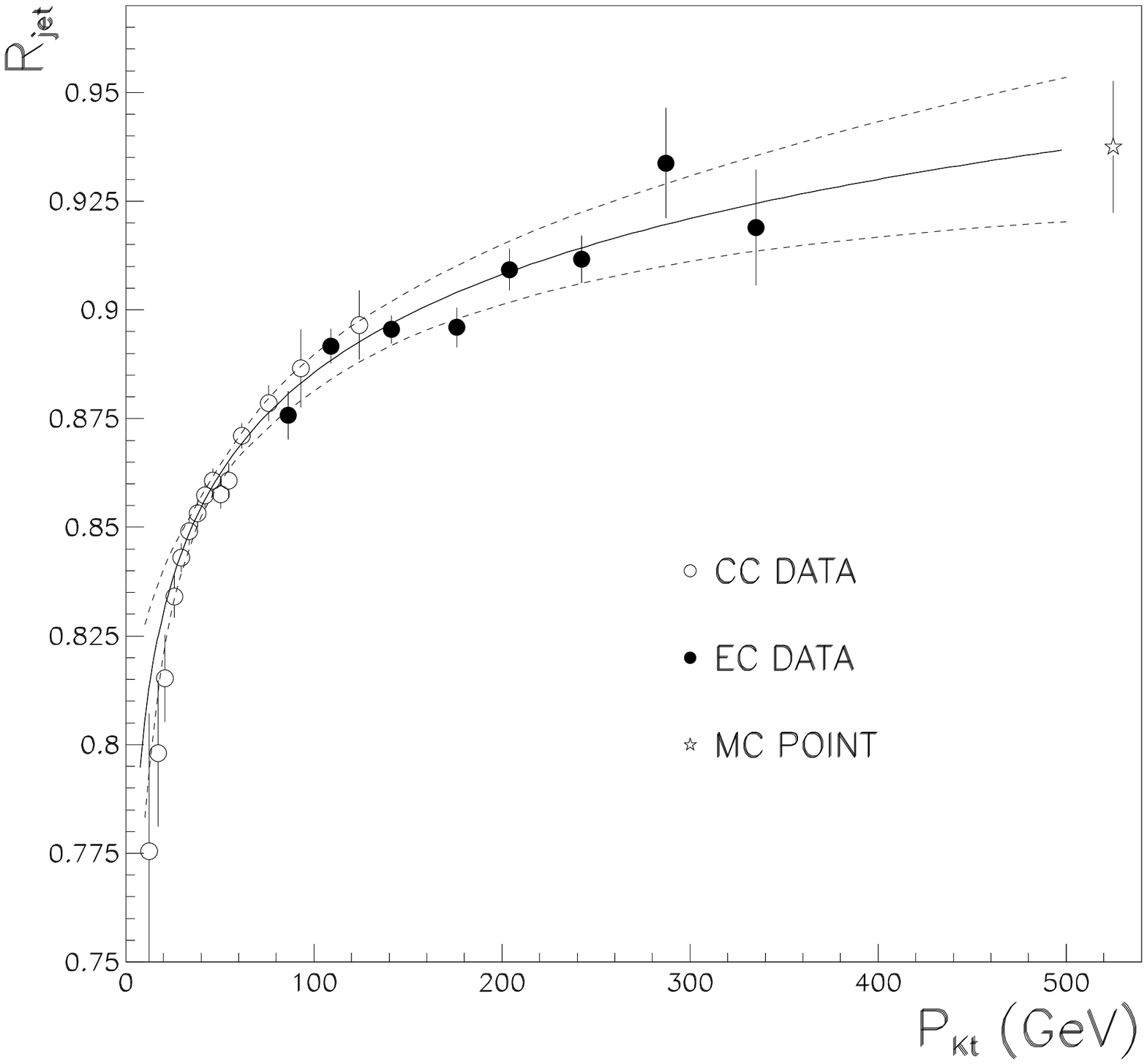,width=3.05in}}}
\vskip-1cm
\caption{$R_{jet}$ versus \KT jet momentum. The solid lines
        are the fit and the dashed band the error of the fit.
        (The three lowest points have nearly fully correlated
        uncertainties and are excluded from the fit.)}
  \label{fig:mom_dep}
\end{minipage}
\hspace*{2mm}
\vskip-0.5cm
\end{figure}

\subsubsection{Tests of the Method}\label{sec:mc}
The accuracy of the $K_{T}$ jet momentum scale correction was
verified using a {\sc herwig} $\gamma$-jet sample and a fast
version ({\sc showerlib})~\cite{showerlib} of the Run~I detector
simulation using {\sc geant}~\cite{geant}. A Monte Carlo jet
momentum scale was derived and the corrected jet momentum compared
directly to the momentum of the associated particle jet.
Figure~\ref{fig:mcclose} shows the ratio of calorimeter and
particle jet momentum before and after the jet scale correction in
the CC. The vertical bars are statistical errors. Systematic errors
(not shown) are of the order of 0.01--0.02. After the jet
correction is applied, the ratio versus particle jet $p_{T}$ is
consistent with unity within the total uncertainty.

\begin{figure}[htb]
\vskip-1cm
\begin{minipage}[t]{3.05in}
{\centerline{\psfig{figure=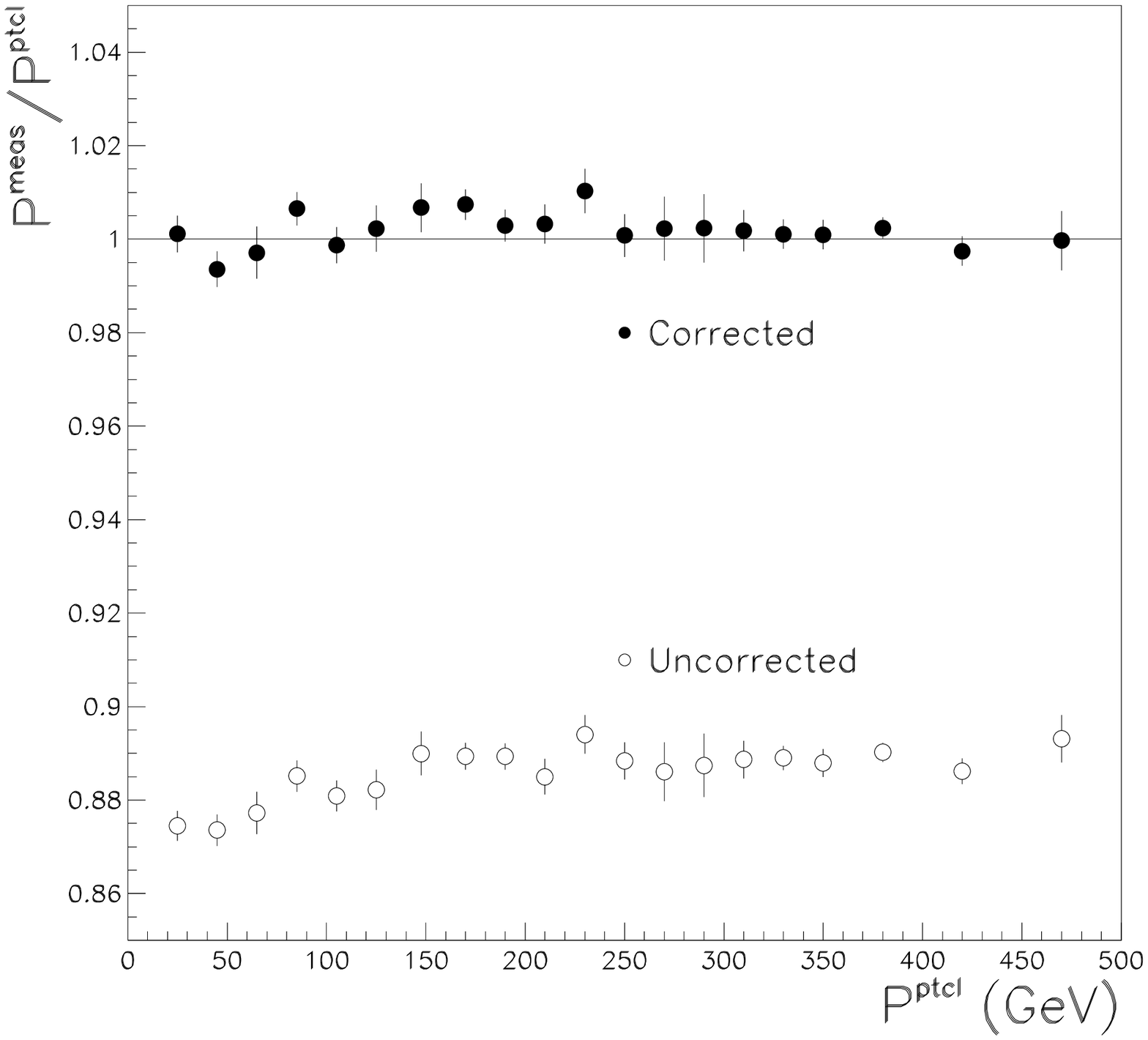,width=3.05in}}}
\vskip-1cm
\caption{Monte Carlo verification test.
        The vertical bars are
statistical errors. Systematic errors (not shown) are of the order
of 0.01--0.02. The corrected
        $p_{jet}^{meas}/p_{jet}^{ptcl}$ ratio is consistent with
        unity within errors.}
  \label{fig:mcclose}
\end{minipage}
\hspace*{2mm}
\vskip-0.5cm
\end{figure}

\subsubsection{Summary}\label{sec:conc}

\D0 improved the method introduced in Ref.~\cite{jes_nim}
for estimating the effects of underlying event, uranium noise,
pile-up, and additional \ppbar\ interactions. The offset correction
is larger for $K_{T}$ jets with $D=1$ than for cone jets with
$R=0.7$ by approximately 20--30\%. The uncertainty ($\sim$0.1~GeV
for underlying event, and $\sim$0.2~GeV for the second offset term
in the CC), however, is slightly smaller. A $K_{T}$ ($D=1$)
algorithm reconstructs more energy from uranium noise, pile-up,
underlying event, and multiple \ppbar\ interactions than a
cone algorithm ($R=0.7$). The accuracy of the associated
corrections are, however, on the same order. The missing $E_{T}$
projection fraction method is well suited to calibrate $K_{T}$
jets~\cite{anna}. The uncertainty in $R_{jet}$ for $K_{T}$ and cone
jets is about the same: (0.5--1.6\%) for jet $p_{T}$ = 50--450~GeV in
the CC.

Overall, it may be possible to determine the jet momentum scale more
accurately for $K_{T}$ jets than the energy scale for cone
jets, given the absence of a cone boundary in the former. The
difference in precision could be large in the low $p_{T}$ and high
pseudorapidity range, where the cone showering correction is
larger and more inaccurately determined. (The showering correction
uncertainty contributes 1--3\%~\cite{anna} to the total error for
$R=0.7$ cone jets.)

\subsection{Jet Momentum Resolutions of $K_{T}$ Jets}
\label{KT_MOMRES}

One of the largest sources of uncertainty in jet measurements
(besides the jet momentum scale) is the effect of a finite
calorimeter jet momentum resolution. {\it A priori}, due to the
absence of cone boundaries, $K_{T}$ jets should be affected little
by jet-to-jet fluctuations in the shower development.  The jets will, of
course, still be subjected to the effects of hadronization.

We compared jet energy resolutions for
cone jets ($R=0.7$) and momentum resolutions for
$K_{T}$ jets ($D=1$) derived from
a \D0 Monte Carlo simulation using the {\sc herwig} event generator plus the
{\sc geant} simulation of the \D0 detector (Run~I). The test was performed
for an inclusive jet sample with average $p_{T}$ = 60~GeV and 80~GeV in
$|\eta|<0.5$. Within statistical errors, $\sigma_{p_{T}}/p_{T}$ for
\KT jets and $\sigma_{E_{T}}/E_{T}$ for cone jets
are the same:
0.109 $\pm$ 0.009 and 0.105 $\pm$ 0.006 for $K_{T}$ ($D=1$) and cone
($R=0.7$) jets at 60~GeV, and 0.10 $\pm$ 0.01 for both at 80~GeV.
Preliminary measurements
of $K_{T}$ jet
momentum resolutions and cone jet energy resolutions using Run~I
collider data support the previous statement.
Note, however, that resolutions depend on the algorithm parameters
$R$ and $D$. Resolution studies for
different (smaller) $R$ and $D$ parameters should be performed, as well as
for different type of samples, for example quark or gluon enriched samples.
These studies will make more clear how energy/momentum resolutions
compare for cone and $K_{T}$ jets.

\subsection{Testing QCD with the $K_{T}$ Jet Algorithm}
\label{KT_test}

\subsubsection{Jet Structure}
\label{subs:structure}
The subjet multiplicity is a natural observable of a $K_{T}$
jet~\cite{sey_sub1,sey_sub2}. Subjets are defined by re-running the
$K_{T}$ algorithm starting with a list of preclusters in a jet.
Pairs of objects with the smallest $d_{ij}$ are merged successively
until all remaining $d_{ij}$ are larger than
$y_{cut} E_{T}^{2}(jet)$, where 0
$<y_{cut}<$ 1 is a resolution parameter. The resolved objects are
called subjets, and the number of subjets within the jet is the
subjet multiplicity $M$. For $y_{cut}=1$, every jet consists of a
single subjet ($M$ = 1). As $y_{cut}$ decreases, the subjet
multiplicity increases until every precluster becomes resolved as a
separate subjet. At this level of detail the specific preclustering
algorithm used clearly influences the result. A measurement of $M$
for quark and gluon jets is a test of QCD, and may eventually be
used in Run II as a discriminant variable to tag quark jets in the
final state. Fig.~\ref{fig:subjets} shows a preliminary
measurement of $M$ by \D0~\cite{dis99}, using Run I data ($K_{T}$
algorithm with $D=0.5$ and $y_{cut}=0.001$). The ratio
$R=\frac{\langle M_{g} \rangle - 1}{\langle M_{q} \rangle - 1}$ is
$1.91 \pm 0.04(stat) \pm 0.23(sys)$. It is well described by the
{\sc herwig} Monte Carlo, and illustrates the fact that gluons
radiate more than quarks.

\begin{figure}[htb]
\vskip-1cm
\begin{minipage}[t]{3.05in}
{\centerline{\psfig{figure=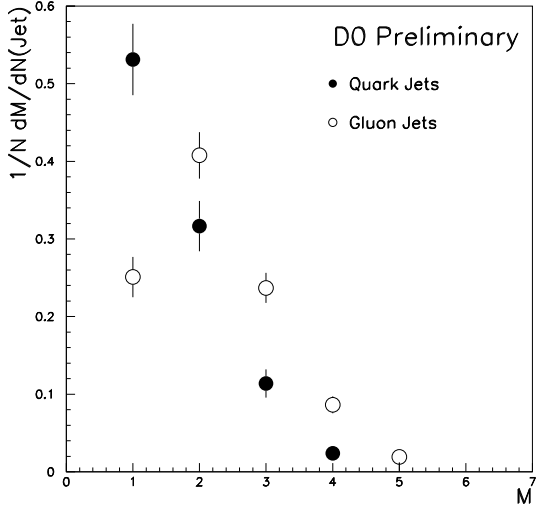,width=3.05in}}}
\vskip-1cm
\caption{Subjet multiplicity for quark and gluon jets at \D0.}
  \label{fig:subjets}
\end{minipage}
\hspace*{2mm}
\vskip-0.5cm
\end{figure}

\subsubsection{Jet Production}

Jet cross section measurements have been extensively used by both
Fermilab Tevatron collaborations during Run I to test perturbative (NLO) QCD
predictions, to test the available parton distribution functions at
the $x$ and $Q^{2}$ ranges covered by the Tevatron, and to
search for quark
compositeness~\cite{incd0,inccdf,angd0,angcdf,dijetd0,dicdf,ratiod0,forwardd0,tripled0,triplecdf}.
The higher center-of-mass energy and the larger data sample will
allow the Tevatron experiments to extend the energy reach and
precision of jet cross sections in Run~II. The largest source of
uncertainty in a jet cross section measurement is the jet energy
(or momentum) scale. As an example, a 1\% uncertainty in the jet
energy calibration translates into a 5--6\% (10--15\%) uncertainty
in the $|\eta|<0.5$ inclusive jet cross section at 100~GeV
(450~GeV). As a function of $\eta$, the jet cross section falls
more quickly with transverse energy, and the cross section error is
even larger.

The $K_{T}$ jet algorithm may provide experimental advantages for
jet production measurements. At \D0, the jet scale uncertainty for
cone jets in the high $E_{T}$ range is dominated by the
contributions from the response and out-of-cone showering
corrections. In Run~II, the availability of more high $E_{T}$
photon data and a more accurate determination of the position of
the interaction vertex promise a reduction in the response
uncertainty. Furthermore, the absence of out-of-cone showering
losses in \KT jets will likely lead to improved jet cross section
measurements in the forward $\eta$ regions. Most of the Run~I cross
section results by CDF and \D0 use jet energy measurements
restricted to central regions ($|\eta|<1$). A couple of exceptions
to the rule are the \D0 measurements of the pseudorapidity
dependence of the jet cross section~\cite{forwardd0}, and the test
of BFKL dynamics in dijet cross sections at large pseudorapidity
intervals~\cite{bfkl}.

\subsubsection{Event Shapes}

Event shape variables in $e^{+}e^{-}$ and $ep$ interactions have
attracted considerable interest over the last few
years~\cite{gardi,anto,sterman}. Little attention has been paid to
measurements or calculations of event shape variables at hadron
colliders. An important example is thrust which is defined as:
\begin{equation}
T=\max_{\hat{n}}\frac{\sum_i |\vec{p_{i}}\cdot \hat{n}|}
{\sum_i |\vec{p_{i}}|} \, , \label{thrust1}
\end{equation}
where the sum is over all parton, or particle momenta.

A LO jet rate calculation with two partons in the final state
yields $T=1$. A NLO calculation, with three partons in the final
state would produce a deviation from $T=1$ (LO in thrust). A NNLO
prediction with four partons in the final state would then give a
NLO prediction of thrust. At all orders, thrust would take values
from 0.5 to unity. In other words, thrust measures the
pencil-likeness of the event: $T \rightarrow 1$ for back-to-back
events, and $T<1$ as more radiation is present. The low scales
introduced by soft and collinear emission in events with $T\lsim 1$
could be the reason for the observed discrepancy between LO and NLO
calculations and experimental $e^{+}e^{-}$ data~\cite{gardi}.
Resummation of higher-order perturbative terms could lead to a better
understanding of the problem.

The simplest measurements of thrust we can perform are the thrust
distributions in jet events, changing the definition of thrust to
sum over all the jets in the event. In order to be able to resum
logarithms of the jet resolution scale, jets must be defined using
an algorithm such as the $K_{T}$ algorithm~\cite{subsey}. The
contribution of the underlying event, and multiple \ppbar\
interactions in hadron colliders, introduce an experimental
difficulty not present in lepton colliders. It is possible,
however, to minimize these systematics by choosing carefully the
variable to measure.

We can also define transverse thrust, $T_{T}$, by replacing
particle momenta by transverse momenta in Eq.~\ref{thrust1}.
$T_{T}$ is Lorentz
invariant for boosts along the beam axis, an advantage in the case
of hadron colliders.

Figs.~\ref{fig:vero3}--\ref{fig:vero1} show the difference between
$T_{T}$ calculated from particle-level jets (reconstructed from
final-state hadrons) and $T_{T}$ from calorimeter-level jets
(reconstructed from cells). {\sc herwig} was used as the generator,
and {\sc showerlib}~\cite{showerlib} (a fast version of {\sc
geant}) simulated the Run~I detector. In all cases jets are
reconstructed with the $K_{T}$ jet algorithm ($D=1$).
Fig.~\ref{fig:vero3} shows a $T_{T}$ distribution for events with
$H_{T}$~=~90--150~GeV, where $H_{T}$ is the scalar sum $p_{T}$ of all
jets above 8~GeV. $H_{T}$ was chosen instead of $Q^{2}$ as an
estimator of the hard scattering energy scale of the event. All
jets with $p_{T}>8$~GeV contribute to $T_{T}$. The full circles are
the particle-level or ``true'' distribution. The triangles are the
distribution as seen in the calorimeter in an ideal environment
with no offset (underlying event, multiple \ppbar\ interactions,
pile-up, or noise). The open circles are a calorimeter-level
distribution which includes a random collider crossing event at a
luminosity of $5\times 10^{30}$cm$^{-1}$sec$^{-1}$. While the
effect of calorimeter momentum response, resolution, and showering
is minimal, the offset distorts the distribution to a large extent.

\begin{figure}[htb]
\vskip-1cm
\begin{minipage}[t]{3.05in}
{\centerline{\psfig{figure=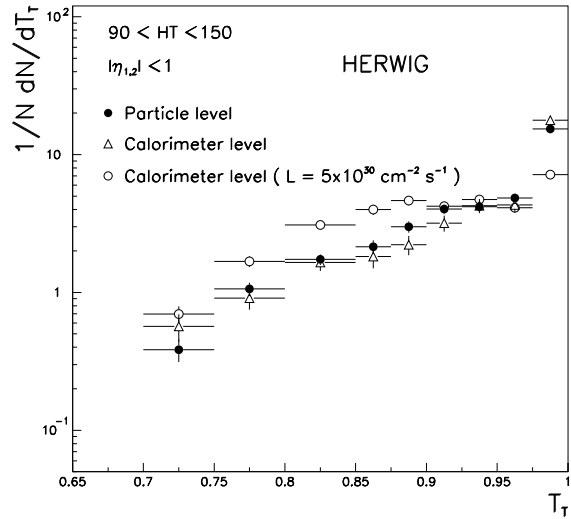,width=3.05in}}}
\vskip-1cm
\caption{All jets with $p_{T}>8$~GeV contribute to $T_{T}$.
The full circles are the particle-level or ``true'' distribution.
The triangles are the distribution as seen in the calorimeter in an
ideal environment with no offset (underlying event, multiple
\ppbar\ interactions, pile-up, or noise). The open circles are a
calorimeter-level distribution which includes a random crossing
collider event at a luminosity of $5\times
10^{30}$cm$^{-2}$sec$^{-1}$.}
  \label{fig:vero3}
\end{minipage}
\hspace*{2mm}
 \vskip-0.5cm
\end{figure}

\begin{figure}[htb]
\vskip-1cm
\begin{minipage}[t]{3.05in}
{\centerline{\psfig{figure=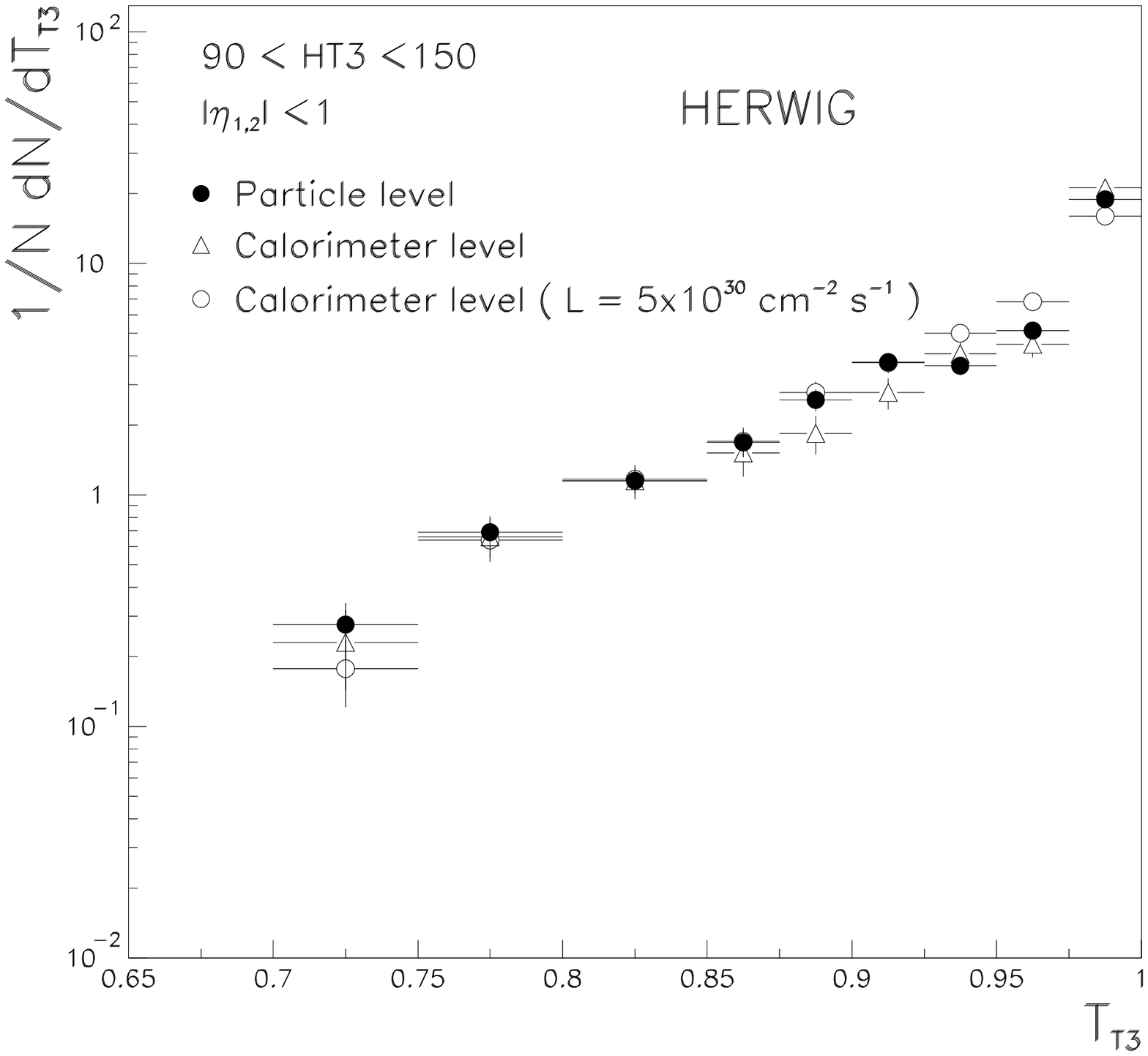,width=3.05in}}}
\vskip-1cm
\caption{Same as Fig.~\ref{fig:vero3} but only the three leading jets
contribute to $T_{T}$, now $T_{T3}$. $H_{T3}$ is the scalar sum $p_{T}$
of the three leading jets in the event.}
  \label{fig:vero2}
\end{minipage}
\hspace*{2mm}
\vskip-0.5cm
\end{figure}

\begin{figure}[htb]
\vskip-1cm
\begin{minipage}[t]{3.05in}
{\centerline{\psfig{figure=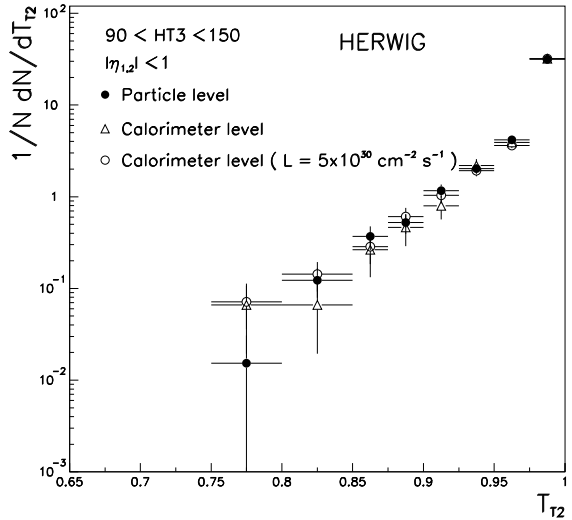,width=3.05in}}}
\vskip-1cm
\caption{Same as Fig.~\ref{fig:vero3} but only the two leading jets
contribute to $T_{T}$, now $T_{T2}$. $H_{T3}$ is the scalar sum $p_{T}$
of the three leading jets in the event.}
  \label{fig:vero1}
\end{minipage}
\hspace*{2mm}
\vskip-0.5cm
\end{figure}

In Fig.~\ref{fig:vero2}, the thrust definition was modified to
allow only the three leading jets (above 8~GeV) to contribute to
the thrust ($T_{T3}$) and to $H_{T}$ (now $H_{T3}$). The difference
between the true and the fully-simulated calorimeter distribution
is now much smaller. Finally, in Fig.~\ref{fig:vero1}, only the
two leading jets contribute to the thrust ($T_{T2}$) for events
with $H_{T3}$~=~90--150~GeV. Now the calorimeter distribution is
even closer to the true distribution. Although $T_{T3}$ and
$T_{T2}$ are not calculated from all final-state particles (to
reduce contamination), they implicitly include the information
about the whole radiation pattern through the $p_{T}$ and
$\eta-\phi$ position of the first few leading jets.

Event shape variables, like a modified version of thrust, can be
studied with precision at the Tevatron. The use of the $K_{T}$
algorithm, infrared safe at all orders in perturbation theory,
provides a test of the newly available hadronic three jet
production calculations at NLO~\cite{nlo3jet,giepri}. In the QCD
calculation of the thrust variables defined in this section, there
are no large logarithms in the $T \rightarrow 1$ limit. Then, it is
neither possible nor necessary to resum them. However, if we
redefine thrust in terms of subjets or tracks, the measurement is
more interesting and resummation becomes an issue~\cite{seypri}.
The availability of the contributions of higher-order terms through
a resummation calculation would be desirable, in that case, to
improve the understanding of the range $T \lsim 1$. In Run II, both
the CDF and the \D0 detectors will have upgraded tracking systems.
This will allow both experiments to implement improved techniques
for the identification of hadrons using both the calorimeter and the
tracking detectors.

The $H_{T}$ dependence of $\langle 1-T \rangle$, in the range where
resummation and hadronization effects are small, could also provide
a measurement of $\alpha_{s}$.

\section{Conclusions}
Jet algorithms present a challenge to experimentalists and
theorists alike.  Although everyone ``knows a jet when they see
it,'' precise definitions are elusive and detailed.  The jet
working group has attempted to provide guidelines and
recommendations for jet algorithm development.  The end product of
the year-long effort has been standardized jet cone and \KT
algorithms, and the recommendation to use 4-vector, E-scheme
kinematic variables. A legacy algorithm or ILCA has been suggested
which will bridge the gap between past results and improved
theoretical calculations.  This document has addressed
concerns about the use and calibration of \KT jets.

We strongly recommend that both CDF and \D0 adopt standard
algorithms for Run II.  Since continued development is probably
inevitable, we encourage continued dialogue. The usefulness of
standardized algorithms, which can replicate past results and meet
experimental and theoretical requirements, makes continued
coordination well worth the effort.

\section{Acknowledgments}
We would like to thank Stefano Catani, Dave Soper and the other
members of the Les Houches QCD working group for stimulating
discussions which lead to the final definition of ILCA. We are also
pleased to thank Rick Field, David Stuart, and numerous members of
the CDF and D\O\ Collaborations for many helpful discussions.
Financial support by the Department of Energy and National Science
Foundation (USA) and CONICET and UBACyT (Argentina) is gratefully
acknowledged. Stephen D. Ellis would also like to thank the
University of Washington Office of Research for partial support and
Fermilab for a Fermilab Frontier Fellowship through the Theoretical
Physics Department.

\end{document}